\RequirePackage[2020-02-02]{latexrelease} 
\documentclass[twocolumn]{aastex63}
\shorttitle{AMUSE$^2$ III: dusty star-forming galaxies within the Slug ELAN}
\shortauthors{Chen et al.}
\graphicspath{{./}{figures/}}
\usepackage{amsmath,amssymb}
\DeclareUnicodeCharacter{0308}{HERE!HERE!}

\begin{document}

\title{A Multiwavelength Study of ELAN Environments (AMUSE$^2$): Detection of a dusty star-forming galaxy within the enormous Lyman $\alpha$ nebula at $z=2.3$ sheds light on its origin}

\correspondingauthor{Chian-Chou Chen (TC)}
\email{ccchen@asiaa.sinica.edu.tw}

\author[0000-0002-3805-0789]{Chian-Chou Chen}
\affiliation{Academia Sinica Institute of Astronomy and Astrophysics (ASIAA), No. 1, Sec. 4, Roosevelt Rd., Taipei 10617, Taiwan}

\author{Fabrizio Arrigoni Battaia}
\affiliation{Max-Planck-Institut fur Astrophysik, Karl-Schwarzschild-Str 1, D-85748 Garching bei M\"{u}̈nchen, Germany}

\author{Bjorn H. C. Emonts}
\affiliation{National Radio Astronomy Observatory, 520 Edgemont Road, Charlottesville, VA 22903, USA}
%%\collaboration{1}{(AAS Journals Data Scientists collaboration)}

%\author{Joseph F. Hennawi}
%\affiliation{Department of Physics, University of California, Santa Barbara, CA 93106, USA}
%\affiliation{Leiden Observatory, Leiden University, P.O. Box 9513, NL-2300 RA Leiden, the Netherlands}

\author{Matthew D. Lehnert}
\affiliation{Sorbonne Universit\'{e}, CNRS UMR 7095, Institut d’Astrophysique de Paris, 98bis bvd Arago, F-75014, Paris, France}

\author{J. Xavier Prochaska}
\affiliation{Department of Astronomy and Astrophysics, UCO/Lick Observatory, University of California, 1156 High Street, Santa Cruz, CA 95064, USA}
\affiliation{Kavli Institute for the Physics and Mathematics of the Universe (Kavli IPMU), WPI, The University of Tokyo Institutes for Advanced Study (UTIAS), TheUniversity of Tokyo, Kashiwa 277-8583, Japan}
%%  possible co-As : Joe Hennawai, X Prochaska, and anyone on Bjorn's proposal.

%\affiliation{Leiden University}
%\affiliation{AAS Journals Associate Editor-in-Chief}
%\nocollaboration{1}

%%\author{Amy Hendrickson}
%%\altaffiliation{AASTeX v6+ programmer}
%%\affiliation{TeXnology Inc.}

% \collaboration{1}{(LaTeX collaboration)}

%%\author{Julie Steffen}
%%\affiliation{AAS Director of Publishing}
%%\affiliation{American Astronomical Society \\
%%1667 K Street NW, Suite 800 \\
%%Washington, DC 20006, USA}

%%\author{Scott Chernoff}
%%\affiliation{IOP Publishing, Washington, DC 20005}

%%\nocollaboration{2}



\begin{abstract}

We present ALMA observations on and around the radio-quiet quasar UM287 at $z=2.28$. Together with a companion quasar, UM287 is believed to play a major role in powering the surrounding enormous Ly$\alpha$ nebula (ELAN), dubbed the Slug ELAN, that has an end-to-end size of 450 physical kpc. In addition to the quasars, we detect a new dusty star-forming galaxy (DSFG), dubbed the Slug-DSFG, in 2\,mm continuum with a single emission line consistent with CO(4-3). The Slug-DSFG sits at a projected distance of 100\,kpc south-east from UM287, with a systemic velocity difference of $-360\pm30$\,km\,s$^{-1}$ with respect to UM287, suggesting it being a possible contributor to the powering of the Slug ELAN. With careful modeling of SED and dynamical analyses it is found that the Slug-DSFG and UM287 appear low in both gas fraction and gas-to-dust ratio, suggesting environmental effects due to the host massive halo. In addition, our Keck long-slit spectra reveal significant Ly$\alpha$ emissions from the Slug-DSFG, as well as a Ly$\alpha$ tail that starts at the location and velocity of the Slug-DSFG and extends towards the south, with a projected length of about 100\,kpc. Supported by various analytical estimates we propose that the Ly$\alpha$ tail is a result of the Slug-DSFG experiencing ram pressure stripping. The gas mass stripped is estimated to be about 10$^9$\,M$_\odot$, contributing to the dense warm/cool gas reservoir that is believed to help power the exceptional Ly$\alpha$ luminosity.
%Finally, the relatively low level of the estimated stripped mass suggests that the main cause for the lower gas fraction and gas-to-dust ratio observed in the Slug-DSFG is instead more likely to be strangulation.

\end{abstract}

\keywords{Galaxies:high-redshift --- Galaxies:halos --- Galaxies:clusters:general --- (Cosmology:)large-scale structure of Universe --- Galaxies: evolution}


\section{Introduction} \label{sec:intro}
Studies of stellar mass functions in the nearby Universe have found that the stellar mass budget is dominated by the early type galaxies, in particular at the massive end ($\gtrsim10^{10}$\,M$_\odot$; \citealt{Bell:2003aa,Kelvin:2014aa,Moffett:2016aa}). These early type galaxies are found to predominantly locate in dense environments and have ceased to form new stars for at least the last several gigayears \citep{Dressler:1980aa,Thomas:2010aa}. These results suggest that the formation of the majority of the stellar masses of the massive galaxies happened at high redshifts, and within very active and dusty environments \citep{Lilly:1999lr}. Indeed, one working hypothesis is that the progenitors of massive local ellipticals involve $z\gtrsim1$ dusty star-forming galaxies (DSFGs) that are undergoing extensive star formation and black hole accretion via dynamically dissipative processes such as mergers (e.g., \citealt{Narayanan:2010aa,Chen:2015aa}), which subsequently transitioned to a quasar phase. The strong feedback from massive stars and black hole accretion quenches star formation and the growth of the black hole, turning the remains into compact quiescent galaxies. By merging these quiescent galaxies, eventually they become the local massive early-type galaxies (e.g., \citealt{Hopkins:2006aa,Alexander:2012aa,Toft:2014aa}).

While seemingly attractive, various places along this hypothetical formation and evolution path of the local massive ellipticals remain unclear. One such place is the transition between DSFGs and quasars. The intimate link between these two populations has been suggested from population studies. First, the evolution of black hole accretion rate density matches well in shape with that of star-formation rate density, both peaking at $z\sim2$, during which the dominant star-forming population is the DSFG \citep{Hopkins:2007aa,Delvecchio:2014aa,Madau:2014aa}. Second, from the point of view of large-scale spatial distribution, statistical measurements of auto-correlation functions of each population have inferred similar correlation lengths, i.e. comparable halo masses, within a wide redshift range of $z\sim1-4$ \citep{Myers:2006aa,Porciani:2006aa,Shen:2007aa,Hickox:2012kk,Eftekharzadeh:2015aa,Chen:2016ac,Wilkinson:2017aa,An:2019aa,Lim:2020aa,Stach:2021aa}. A certain level of overlap of the two populations in both space and time can find support from cross-correlation measurements \citep{Wang:2015aa}, however, due to low space density of both populations, detailed assessments of their environmental connections, in particular on halo scales, have been difficult. 

One way to tackle this issue of low space densities is to conduct targeted far-infrared or submillimeter observations around selected samples of $z\gtrsim1$ quasars, aiming to uncover and characterise physically associated DSFGs around central quasars. Indeed, via submillimeter number counts measured from single-dish surveys, populations of DSFGs have been statistically found on Mpc scales around various quasar samples such as high-redshift radio galaxies (HzRGs) and AGN selected from the Wide-field Infrared Survey Explorer({\it WISE}) all-sky infrared survey \citep{Stevens:2003aa,Rigby:2014aa,Dannerbauer:2014aa,Zeballos:2018aa}, corroborating the co-existing nature on halo scales between DSFGs and quasars. On the other hand, interferometric observations have allowed discovery and characterisations of DSFGs around quasar samples on $\sim$100\,kpc scales, often finding evidence of galaxy mergers that in principle helped trigger the vigorous activities in these regions \citep{Ivison:2012aa,Emonts:2015ab,Gullberg:2016ab,Trakhtenbrot:2017aa,Decarli:2018aa,Hill:2019aa}. However, these studies were often focused on individual cases or small samples, thus extending them into different kinds of quasars are needed to fully capture the environmental relations between DSFGs and quasars.

To do so, we have initiated A Multiwavelength Study of ELAN Environments (AMUSE$^2$) campaign, with the primary goal to study the multi-phase environment of quasars that host enormous Ly$\alpha$ nebulae (ELANe; \citealt{Arrigoni-Battaia:2018aa, Arrigoni-Battaia:2021aa, Nowotka:2021aa}), particularly from the far-infrared and submillimeter point of view. Only recently discovered, ELANe represent an extreme form of Ly$\alpha$ emission at $z\sim2-3$, which typically extend over $>100$\,kpc in physical scales with a surface brightness level of SB$_{\rm Ly\alpha} \sim10^{-17}$~erg~s$^{-1}$~cm$^{-2}$~arcsec$^{-2}$ \citep{Cantalupo:2014aa,Hennawi:2015aa,Cai:2017aa,Cai:2018aa,Cai:2019aa,Arrigoni-Battaia:2018ab}. 

In order to explain the extreme Ly$\alpha$ luminosity and the measured line ratios between Ly$\alpha$ and CIV and HeII, detailed spectral synthesis modelling has suggested that these nebulae, powered by multiple sources, represent a large amount of dense ($n\sim1$\,cm$^{-3}$) and cool ($T\sim10^4$\,K) gas with masses about 10$^{10-11}$\,M$_\odot$ \citep{Arrigoni-Battaia:2015aa,Hennawi:2015aa,Cai:2017aa}. Molecular gas measurements based on CO(1-0) on one of the ELAN, the MAMMOTH-1 ELAN, have indeed shown a comparable amount of cold gas in the circumgalactic medium (CGM) around the powering quasars \citep{Emonts:2019aa}. This large cool and cold gas reservoir could originate from cold accretion from the IGM or stripped/expelled gas from the infalling galaxies. Evidence also suggests the possibilities of undetected obscured sources to partly contribute to ELAN via softer ionizing radiations from star formation \citep{Arrigoni-Battaia:2018ab}.

In this paper we focus on a proto-typical ELAN, the Slug ELAN, which was first discovered by \citet{Cantalupo:2014aa} using Keck narrow-band imaging. With an end-to-end size of about 450\,kpc, the Slug ELAN remains one of the most luminous and the largest in Ly$\alpha$ among all the Ly$\alpha$ nebulae discovered so far (see \citet{Ouchi:2020aa} for a review). Follow-up observations have painted a scenario where the nebula consists of multiple gaseous components that are distinct in kinematics. {\ There appears to be three main kinematic structures; If referencing to UM287, the first is found to be at about $400$\,km\,s$^{-1}$, dubbed region c in \citet{Cantalupo:2019aa}, consisting of compact sources `C' and `D' reported by \citet{Leibler:2018aa} as well as the He\,{\sc ii} extended emission. The second component is at about $-500$\,km\,s$^{-1}$, corresponding to the 'bright tail' of the Ly$\alpha$ nebula. And finally the third component of the Ly$\alpha$ nebula can be found at about $-300$\,km\,s$^{-1}$, which was dubbed region 1 by \citet{Leibler:2018aa}.} Their origins are still in active discussion \citep{Martin:2015aa,Arrigoni-Battaia:2015aa,Leibler:2018aa,Cantalupo:2019aa}. Here we complement the existing datasets from the submillimeter point of view, aiming to address whether there are DSFGs hidden from optical observations, and if yes what are the roles they play for the presence of the Slug ELAN.

Throughout this paper we adopt the AB magnitude system \citep{Oke:1983aa}, and we assume the {\it Planck} cosmology: H$_0 =$\,67.8\,km\,s$^{-1}$ Mpc$^{-1}$, $\Omega_M = $\,0.31, and $\Omega_\Lambda =$\,0.69 \citep{Planck-Collaboration:2014aa}. At the redshift of the Slug ELAN, 2.28, the angular distance scale is 8.4\,kpc for 1$''$.

\section{Observations and data} \label{sec:obs}
\subsection{ALMA} \label{sec:alma2018}

The 12-meter array observations were split into two execution blocks (EBs), both of which were carried out in mid-January 2019. During that period of time the baselines had a range from 15.1m to 313.7m, the mean precipitation water vapor was 3.6\,mm, and at least 44 antenna were used to collect the data. The four science spectral windows were tuned to a band 4 standard FDM with 1920 channels, two of which were for continuum, and each of the remaining two was focused on redshifted CO(4-3) and [CI] $^{3}$P$_{1}$\,$-$\,$^{3}$P$_{0}$ (hereafter [CI](1-0)), respectively. The primary beam full-width-half-maximum (FWHM) for band 4 with the 12-meter array is about 45$''$ (380\,kpc). The amplitude (flux) and bandpass calibrator was J0006-0623, and J0108+0135 was used for phase monitoring. {\ The total on-source time given by the QA2 report is summed to 2.2 hours.}

We re-imaged the data to produce the optimal results. To do that, the raw visibilities were calibrated by running the corresponding scriptForPI.py scripts under {\sc casa} version 5.4.0-68, meaning that we adopted the calibrations performed by the QA2 team members, who also noted that additional flagging was done on bad antennas DA55 and DA61 in phase calibrator. We confirm the quality of the reduction by examining the weblog. To make the imaging process more efficiently, the calibrated visibilities were first binned in channels by a factor of 10, resulting in a frequency resolution of 9.77 MHz, corresponding to a velocity resolution of 19-21 km/s depending on the exact frequency. We note that our detected lines are much broader than the binned channel width so this step does not affect any of the results. We then proceeded to continuum subtraction. The channels that contain significant line-emission were identified based on the delivered data cubes and excluded from continuum fitting. The identified continuum channels were used for both continuum generation and subtraction. While we later refer 
to all band-4 continuum as 2\,mm continuum, the continuum for this data set has a nominal frequency at 145.1\,GHz, so 2.07\,mm in wavelength. 

We then used {\sc tclean} to perform inverse Fourier transformation and clean on both the continuum and continuum-subtracted visibilities. The visibilities were transformed to images of 270$\times$270 pixels in x (RA) and y (DEC) axis with a pixel size of 0$\farcs$3 ($\sim$6-8 pixels per synthesized beam). Natural weighting for the baselines was chosen to enhance source detection, however later in the analyses we re-image the cubes with other weighting as needed. To increase the efficiency of clean, we adopted the auto-masking approach in Tclean, by setting the usemask papameter to 'auto-multithresh', the noisethreshold parameter to 4, and the lownoisethreshold parameter to 2.5. The cubes within the masked regions determined by auto-masking were then cleaned down to 2 sigma level (nsigma = 2 in Tclean). Finally, given the same baseline weighting across the whole frequency range, the reduced cubes have different spatial resolution in each frequency channel. To allow more straightforward analyses and understanding of the data we applied smoothing on the reduced cubes by using the CASA routine Imsmooth, setting the kernel to the common resolution, which is normally the largest beam size of the pre-smoothed cube. After this step, the spatial resolution is the same across the image cubes, with a size of $2\farcs6\times2\farcs0$ and a P.A. of $-67$ degrees. The synthesized beam shape for the continuum is similar to that of the cubes, $2\farcs5\times1\farcs9$ and a P.A. of also $-67$ degrees. The final cleaned continuum image has a 1\,$\sigma$ sensitivity of 12.4 $\mu$Jy/beam at 145\,GHz, and the mean sensitivity for the cleaned image cube is about 300\,$\mu$Jy/beam per binned channel (9.77\,MHz). The achieved sensitivities are consistent with the expected values from the sensitivity calculator.

\subsection{Multi-band imaging data} \label{subsec:imaging}
We exploit multiple waveband imaging data that we acquired in the past years for the Slug ELAN field: $B$-band, $V$-band, $J$-band, 450 and 850\,$\mu$m. The $B$-band (1~hour) and $V$-band (10~hours) imaging data were taken with the Keck telescope using the LRIS instrument \citep{Oke:1995aa} on UT 12-13 November 2013. The reduction of the data was performed using standard techniques (see \citealt{Cantalupo:2014aa} for details). The $J$-band imaging observations were carried out using the HAWK-I instrument (\citealt{Casali:2006aa}) on the VLT on UT 17 October 2018 under clear weather (Program ID: 0102.C-0589(D), PI: F.~Vogt).
These data were acquired as 16 60s $J$-band exposures to which it is applied a dithering within a jitter box of $15\arcsec$. The Slug ELAN was placed in the fourth quadrant of the $7\arcmin.5 \times 7\arcmin.5$ HAWK-I field-of-view. We reduced these data as done in paper I, using the standard ESO pipeline version 2.4.3 for HAWK-I\footnote{\url{https://www.eso.org/sci/software/pipelines/hawki/hawki-pipe-recipes.html}}. The flux calibration  and astrometry are obtained using the 2MASS catalogue. The astrometry has an average error of $\sim 0.2\arcsec$. The seeing in the final combined image is $0.7\arcsec$. The submillimeter imaging at 450 and 850\,$\mu$m were taken with the SCUBA-2 camera \citep{Holland:2013lr} on the JCMT, and the procedures for reduction, calibration, as well as part of the 850\,$\mu$m results were presented in \citet{Nowotka:2021aa} (Paper II). The Slug ELAN was located at the center of the daisy pattern used during observations. At this location the 450 and 850\,$\mu$m data have an rms of 11.7 and 1.02~mJy, respectively.

\section{Analyses and Results} \label{sec:results}

%FIGURE 1%
\begin{figure*}[ht!]
	\begin{center}
		\leavevmode
		\includegraphics[scale=0.42]{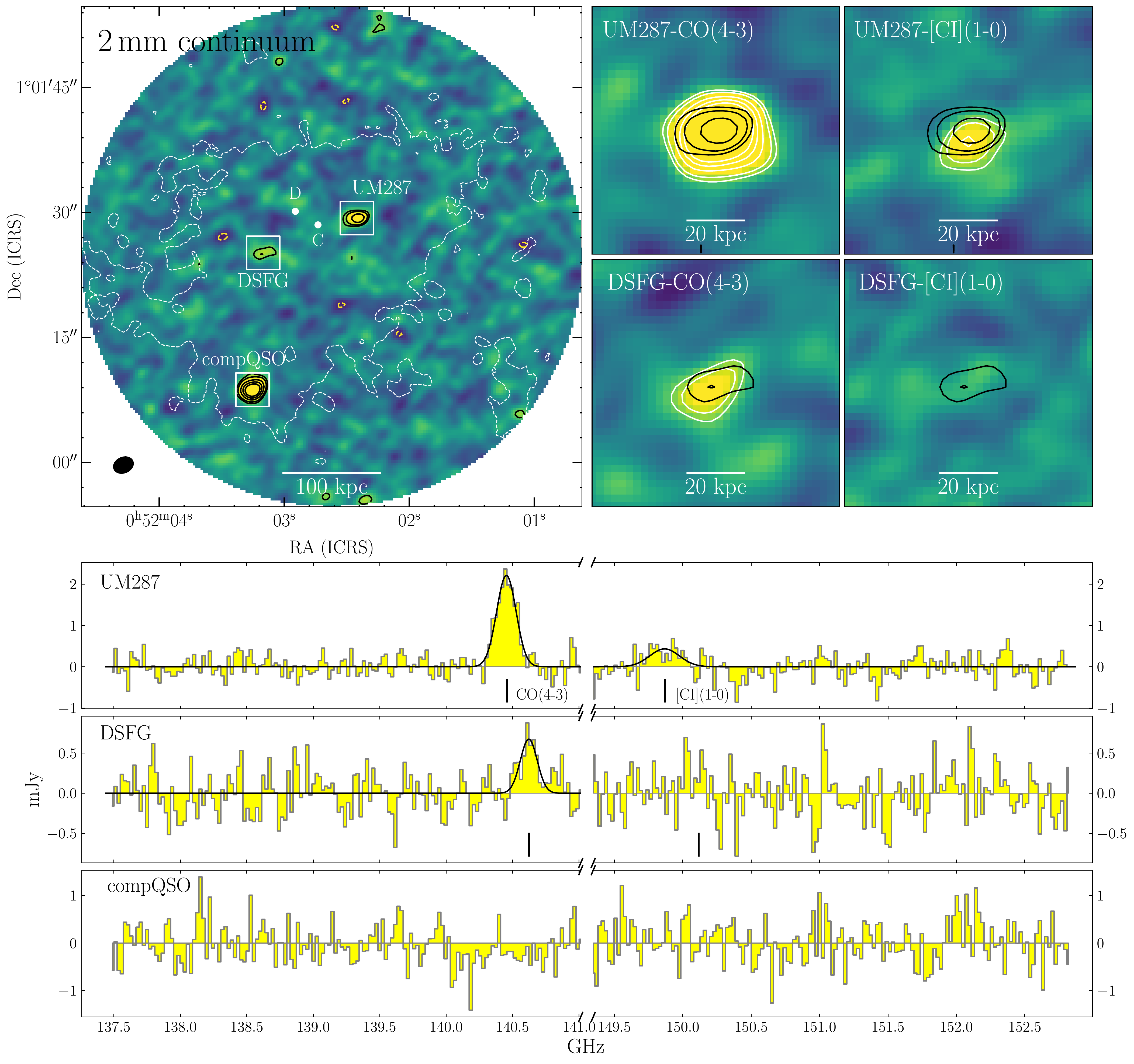}
		\caption{A gallery of our key ALMA results of both 2\,mm continuum and emission lines toward the central regions of the Slug ELAN. {\it Top-left}: 2\,mm continuum image, without primary beam correction, is linearly scaled from -4 to 5\,$\sigma$, and roughly centered on the enormous Ly$\alpha$ nebula. 
		%The red contours 
		The white dashed contours show the Ly$\alpha$ surface brightness isophote of $10^{-18}$\,erg s$^{-1}$ cm$^{-2}$ arcsec$^{-2}$ obtained from Keck narrow-band imaging \citep{Cantalupo:2014aa}. The black solid contours show the positive 2\,mm continuum signals at 3, 4, 6, 8, 10\,$\sigma$, and the dashed yellow contours show the level of -3\,$\sigma$. The three significant continuum detections, enclosed by the Ly$\alpha$ isophote, are marked by white boxes with the corresponding abbreviated source names. {\ The locations of compact sources 'C' and 'D' are shown for a clear overview of this system.}
		The synthesized beam shape and size are shown at the corner in black. {\it Bottom}: Spectra extracted at the three continuum detections respectively using apertures two times of the synthesized beam, with centroids determined through an iterative process detailed in \autoref{subsec:lines}. The line emission that are significantly detected are well fit by single-Gaussian models which are plotted as black curves, and the line identifications are indicated with short vertical segments. Note the spectra are broken in frequency to remove the gap due to the nature of the tuning. {\it Top-right}: Moment zero maps of the CO(4-3) and {\bf [CI](1-0)} lines from UM287 and the DSFG, contoured by white curves with levels same as the continuum. For comparison the continuum is shown in black contours again with the same levels. The physical scale at the UM287 redshift is marked by scale bars as a reference.
		}
		\label{fig:fig1}
	\end{center}
\end{figure*}

\subsection{2 mm continuum} \label{subsec:2mmcont}
The cleaned 2\,mm continuum of the central UM287 region is shown in \autoref{fig:fig1} along with its synthesized beam at the corner. To systematically identify significantly detected sources, we ran {\sc sextractor} on both the original and inverted images and then determined a source-finding threshold above which there is no detection in the inverted image. We subsequently identified three sources that are above such a detection threshold, all of which have a peak signal-to-noise ratio of $>4$. These three sources, UM287, the dusty star-forming galaxy, and the companion quasar, are boxed in \autoref{fig:fig1} with their abbreviated names (UM287, Slug-DSFG, and compQSO). The sky coordinates deduced from {\sc sextractor} were then used as the initial guess for the {\sc casa} routine {\sc imfit}, which was used to fit two-dimensional Gaussian profiles to measure the integrated flux densities and estimate their sizes. We found all three of them unresolved in the current image, as well as in images with higher spatial resolution (slightly under 2$''$) produced with robust weighting in {\sc tclean}. No meaningful size constraints can be derived. Since they are unresolved, we adopt the sky positions and the peak values from {\sc imfit} for the three sources and the primary beam correction was applied to obtain the final flux density measurements. We have also used {\sc uvmultifit} to conduct Gaussian fitting in the visibility space and the results are consistent with those derived from the image-base fitting. The measurements are provided in \autoref{tab:table1} {\ for the three continuum detected sources, and upper limits for source 'C' and 'D' are also provided for completeness.}

\begin{table*}
    \scriptsize    
    \caption{Continuum and Line Measurements\footnote{Uncertainties in 1\,$\sigma$ are given in the parentheses and only the last decimal number is given if the measurements are precise to more than two decimal points. Upper limits are 6\,$\sigma$ based on the line finding code {\sc LineSeeker} (\autoref{subsec:lines}). All flux densities have been corrected for the primary beam effects. In all cases CO represents CO(4-3) and [CI] stands for [CI](1-0)}}             
    \label{tab:table1}      
    \vspace{-4mm}
    \begin{center}
        \begin{tabular}{llllllllllll}
            \hline
            ID & R.A.$_{\text{2mm}}$ & Decl.$_{\text{2mm}}$ & $F_{\text{\text{2mm}}}$ & $z_{\text{CO}}$ & $I_{\text{CO}}$ & FWHM$_{\text{CO}}$ & $L'_{\text{CO}}$\footnote{Line luminosities in units of 10$^9$K\,km/s\,pc$^{\text{2}}$.} & $z_{\text{\ [CI]}}$ & $I_{\text{\ [CI]}}$ & FWHM$_{\text{\ [CI]}}$ & $L'_{\text{\ [CI]}}$$^{\text{b}}$\\

            & [J2000;deg] & [J2000;deg] &[mJy] &  & [Jy\,km/s] & [km/s] &  &  & [Jy\,km/s] & [km/s] & \\
            \hline
            UM287 & 13.010083 &  1.024816 & 0.11(0.01) & 2.2825(1) & 0.87(0.08) & 368(25) & 14(1) & 2.2840(8) & 0.24(0.10) & 513(171) & 3.4(1.5)\\
            Slug-DSFG & 13.013202 &  1.023657 & 0.05(0.01) & 2.2786(4) & 0.23(0.07) & 319(75) & 3.7(1.2)& N/A & $<$0.23 & N/A & $<$3.3\\
            compQSO & 13.013560 & 1.019100 & 0.28(0.02) & N/A & $<$0.23 & N/A & $<$3.7 & N/A & $<$0.23 & N/A & $<$3.3\\
            {\ Source C} & {\ 13.011378} & {\ 1.024590} & {\ $<0.04$} & {\ N/A} & {\ $<$0.23} & {\ N/A} & {\ $<$3.7} & {\ N/A} & {\ $<$0.23} & {\ N/A} & {\ $<$3.3}\\
            {\ Source D} & {\ 13.012132} & {\ 1.025042} & {\ $<0.04$} & {\ N/A} & {\ $<$0.23} & {\ N/A} & {\ $<$3.7} & {\ N/A} & {\ $<$0.23 }& {\ N/A} & {\ $<$3.3}\\
            \hline
        \end{tabular}
    \end{center}
\end{table*}

\subsection{Line emission} \label{subsec:lines}
To systematically search for line emission and quantify their significance, we ran the publicly available code {\sc LineSeeker}, which was first written and used to search for line emission for the ALMA frontier field survey \citep{Gonzalez-Lopez:2017aa}, and was later applied to the data taken for the ALMA Spectroscopic Survey in the Hubble Ultra Deep Field (ASPECS; \citealt{Walter:2016aa, Gonzalez-Lopez:2019aa}). Similar to another publicly available line searching code {\sc mf3d} \citep{Pavesi:2018aa}, {\sc LineSeeker} utilizes a matched filtering kind of technique that combines spectral channels based on Gaussian kernels with a range of widths. The widths are chosen to match lines detected in real observations. {\sc LineSeeker} was chosen for our data over {\sc mf3d} since it was designed specifically for ALMA data cubes that have a similar spectral setting as that of our data. Comparisons between the two codes in fact show that both effectively return consistent results to within 10\% \citep{Gonzalez-Lopez:2019aa}.

With {\sc LineSeeker} we obtained three line detections at $\ge6\,\sigma$ from two sources, with their sky locations matching %to 
those of the two continuum sources, UM287 and the Slug-DSFG. At this significance level there is no equivalent detection from the negative signal, which was at maximum at $\sim$5\,$\sigma$ level. The signal-to-noise of $\ge6$ has also been shown to return robust detections by the ASPECS team. {\ No significant line detection can be found from the companion quasar. As we show later this is likely due to the fact that the millimeter photometry is dominated by synchrotron radiation therefore most of the infrared luminosity is contributed by its super massive black hole, and the sensitivity of our ALMA observations is not sufficiently deep to detect the expected CO or [CI] lines given its relatively faint dust infrared luminosity.} 

The final spectra were obtained in the following steps. Motivated by the results of {\sc LineSeeker} and the fact that sources are not spatially resolved in continuum, we extract the spectra, first centered on the continuum positions, using elliptical apertures with a shape of the synthesized beam. The optimal aperture size to extract the total line flux density was determined by a curve-of-growth analysis, where Gaussian fits were performed on spectra extracted from beam apertures scaled by a range of factors from 0.25 to 2.5. The optimal aperture size was then determined to be the one that larger apertures do not return significantly higher flux densities, and we found for all the lines a factor of 2 of the synthesized beam is optimal. We then produced moment zero maps based on the spectra extracted from the optimal apertures and performed {\sc imfit} to determine their sky positions. We then updated the aperture centroid to the coordinates deduced from {\sc imfit} and iterated the above processes until all relevant measurements converge, including the sky locations, redshifts, and line properties. The results, after applying the primary beam correction, are provided in \autoref{tab:table1} and the final spectra and their associated moment zero maps are plotted in \autoref{fig:fig1}. 

{Like} the continuum, we found that the line emission are not spatially resolved, even if we push the spatial resolution to about 2$''$ in major axis with the robust baseline weighting. In addition, the lines are all well described by a single-Gaussian profile, with reduced $\chi^2$ = 1.0. Lastly, considering the beam sizes and the significance of the detection, the sky locations of the emission lines are consistent with those of continuum, and we therefore only provide the continuum coordinates in \autoref{tab:table1} to avoid confusion. 

Regarding the identification of the lines, we found the two lines detected at the direction of UM287 can only be CO(4-3) and [CI](1-0) assuming both were emitted from the same source.% so at the same redshift. 
While the redshifts independently deduced from the two lines agree with each other to within 2\,$\sigma$ uncertainty, they both significantly deviate from 2.279$\pm$0.001, the redshift typically adopted for UM287 based on its rest-frame optical spectra \citep{McIntosh:1999aa}. Our measured redshift is also in agreement with that measured by \citet{Decarli:2021aa} via CO(3-2), who reported 2.2824$\pm$0.0003. Unlike molecular line spectra, the optical spectra of quasars are complicated by blended Fe{\sc ii} emission lines, and it often requires assuming certain Fe{\sc ii} line templates in order to deduce redshifts. The uncertainty of this template assumption is difficult to quantify and normally not included in the quoted uncertainty. We therefore conclude that the most likely explanation of this offset in redshift is that the uncertainty of the optical redshift was underestimated. Assuming all redshifts agree, given that the one derived from CO(4-3) has the highest precision, we adopt the CO redshift for UM287 for this paper, 
i.e. $z=2.2825\pm0.0001$.

For the single line detection toward the Slug-DSFG, we estimate the probability of it being a random source on the sky and not associated with the UM287 system. We look at the probability from two angles; the line and the continuum. For the line, we estimate the probability by first assuming an empirically motivated redshift distribution model for the DSFG population in general and then integrating the corresponding redshift range from a list of line candidates with the observed frequencies lying within 2$\sigma$ from that of UM287. The redshift distribution of the general DSFG population is modelled as a lognormal function (equation 3 in \citealt{Chen:2016aa}) with a median of 1.85, of which the median is chosen to match the most recent result from \citet{Aravena:2020aa}, who focused on a sample of uniformly selected DSFG that has a similar continuum flux density as the Slug-DSFG. The list of line candidates considered includes CO with upper transition from J=2 to J=9 as well as the two lowest transition level {[CI]} lines. Lines with higher transition would infer a DSFG redshift of $z>6.4$, which given our assumed redshift distribution model would have little impact in the estimates of the random probability. We include the two {[CI]} transition because given the recent survey of {[CI]} on DSFGs (e.g., \citealt{Bothwell:2017aa}) we can not rule out their presence based on the line properties. In the end we find that the probability of detecting a 
non-associated line within the above set condition is about 1\%. 

Now we turn to the continuum detection. The Slug-DSFG has a 2\,mm continuum flux density of 50\,$\mu$Jy, roughly equivalent to 0.2\,mJy at 1.2\,mm assuming a dust emissivity index of 1.8. Based on the 1.2\,mm number counts of the most recent ASPECS results \citep{Gonzalez-Lopez:2020aa} it is expected to find maximum 0.05 (considering upper 3\,$\sigma$ bound) sources given the effective area, the total area {\ that is} sensitive enough to detect a source with a flux density of the Slug-DSFG at $\ge4$\,$\sigma$. In combination, we conclude that the Slug-DSFG is associated with the UM287 system at a confidence level of $>4$\,$\sigma$ ($>3\,$\,$\sigma$ from continuum multiplied by 1\% from the line), and that detected line is CO(4-3), with a velocity difference of --$360\pm30$\,km/s with respect to UM287 assuming that the redshift difference is solely due to peculiar velocity. Note that as seen in \autoref{fig:fig1} there is a 2\,$\sigma$ signal emitting from the frequency range of the expected [CI](1-0) line at the Slug-DSFG location.

For the undetected lines we estimate the upper limits based on our S/N cut of 6\,$\sigma$ from {\sc LineSeeker}. It just so happens that the CO line from the Slug-DSFG has a S/N of 6, we therefore adopt its integrated line flux density as the upper limit for all the undetected lines. This is likely a very conservative estimate since the S/N of 6 cut is based on a blind search without using the prior knowledge. For example by limiting the source redshift on specific lines the upper limits can be lowered. We however opt this approach for simplicity, and this choice does not affect our conclusions. In fact, adopting lower upper limits would strengthen some of the conclusions. The redshift of the companion quasar is assumed to be the same as that of UM287 \citep{Cantalupo:2014aa}{\, and that for both source 'C' and 'D' is assumed to be 2.287 \citep{Leibler:2018aa}. From now on the analyses are focused on the three sources that are detected in the ALMA data.}

\subsection{Multi-band photometry and source properties}\label{subsec:sed}
To estimate the physical properties of the three submilleter detected sources we perform model fitting on their spectral energy distributions (SEDs). We first obtain the photometry for all three sources. For $B$, $V$, and $J$ bands we ran {\sc SExtractor} \citep{Bertin:1996zr} using our imaging data taken from the Keck telescope and the VLT. We adopt the MAG\_AUTO photometric measurements provided by {\sc SExtractor}, which are based on Kron-like elliptical apertures. For non-detections, we simply estimate the 3\,$\sigma$ upper limits by measuring the standard deviation of flux densities of 500 random, source-free positions using a 2$''$ diameter aperture. The results are presented in \autoref{tab:table2}. For the rest of the wavebands we cross match the three sources to the appropriate public catalogs and adopt the provided photometry measurements. We also estimate 3\,$\sigma$ upper limits wherever the imaging sensitivity is publicly available, which is the case for wavebands long-ward of {\it K$_{\rm s}$} band except for the photometry adopted from the ALLWISE catalog, in which 5\,$\sigma$ upper limits are provided. 

\begin{table}
    \begin{center}
        \caption{Multi-band photometry of the three main sources}
        \label{tab:table2} 
       \begin{tabular}{llrrr}
            \hline
            Band/ID & UM287 & compQSO & Slug-DSFG & References\\
            \hline
            FUV & 0.0175(9) & $\cdot\cdot\cdot$ & $\cdot\cdot\cdot$ & (1)\\
            NUV & 0.0249(8) & $\cdot\cdot\cdot$ & $\cdot\cdot\cdot$ & (1)\\
            $u$ & 0.259(4) & $\cdot\cdot\cdot$ & $\cdot\cdot\cdot$ & (2)\\
            $B$ & 0.265(2) & 0.0068(2) & $<$4.0E-05$^b$ & This work\\
            $g$ & 0.397(7) & $\cdot\cdot\cdot$ & $\cdot\cdot\cdot$ & (2)\\
            $V$ & 0.1775(5)$^a$ & 0.0084(1) & $<$3.6E-05$^b$ & This work\\
            $r$ & 0.422(4) & $\cdot\cdot\cdot$ & $\cdot\cdot\cdot$ & (2)\\
            $i$ & 0.45$\pm$0.07 & $\cdot\cdot\cdot$ & $\cdot\cdot\cdot$ & (2)\\
            $z$ & 0.562(9) & $\cdot\cdot\cdot$ & $\cdot\cdot\cdot$ & (2)\\
            $J$ & 0.538(2) & 0.0058(3) & $<$0.001$^b$ & This work\\
            $H$ & 0.52$\pm$0.07 & $\cdot\cdot\cdot$ & $\cdot\cdot\cdot$ & (3)\\
            $K_{\rm s}$ & 0.70$\pm$0.06 & $\cdot\cdot\cdot$ & $\cdot\cdot\cdot$ & (3)\\
            W1 & 0.55$\pm$0.02 & $<$0.04 & $<$0.04 & (4)\\
            W2 & 0.77$\pm$0.03 & $<$0.09 & $<$0.09 & (4)\\
            W3 & 2.7$\pm$0.3 & $<$0.7 & $<$0.7 & (4)\\
            W4 & 6.1$\pm$1.7 & $<$6.9 & $<$6.9 & (4)\\
            450 & $<$35.1 & $<$35.8 & $<$35.3 & This work\\
            850 & $<$3.1 & $<$3.1 & $<$3.1 & This work\\
            2\,mm & 0.11$\pm$0.01 & 0.28$\pm$0.02 & 0.05$\pm$0.01 & This work\\
            3\,mm & $<$0.08 & 0.33$\pm$0.05 & $<$0.08 & (5)\\
            1.4\,GHz & $<$0.3 & 24.2$\pm$0.8 & $<$0.4 & (6)\\
            \hline
        \end{tabular}
        \begin{tabular}{l}
            Note: \\
            All units in mJy. \\
            Uncertainties in parentheses if the measurements are precise \\
            to better than two decimal points. \\
            $^a$: Saturated. \\
            $^b$: 3\,$\sigma$ of 2$''$ diameter aperture on the source-free regions \\ 
            References: \\
            (1) \citet{Bianchi:2017aa}; GALEX all-Sky Catalogue \\
            (2) \citet{Paris:2017aa}; SDSS DR12 QSO catalog \\
            (3) 2MASS all-sky point source catalog \\
            (4) AllWISE Source Catalog \citep{Wright:2010aa} \\
            (5) \citet{Decarli:2021aa} \\
            (6) \citet{Condon:1998aa}; NRAO VLA Sky Survey (NVSS). 
    	\end{tabular}    
	\end{center}
\end{table}

\begin{figure}[ht!]
	\begin{center}
		\leavevmode
		\includegraphics[scale=0.5]{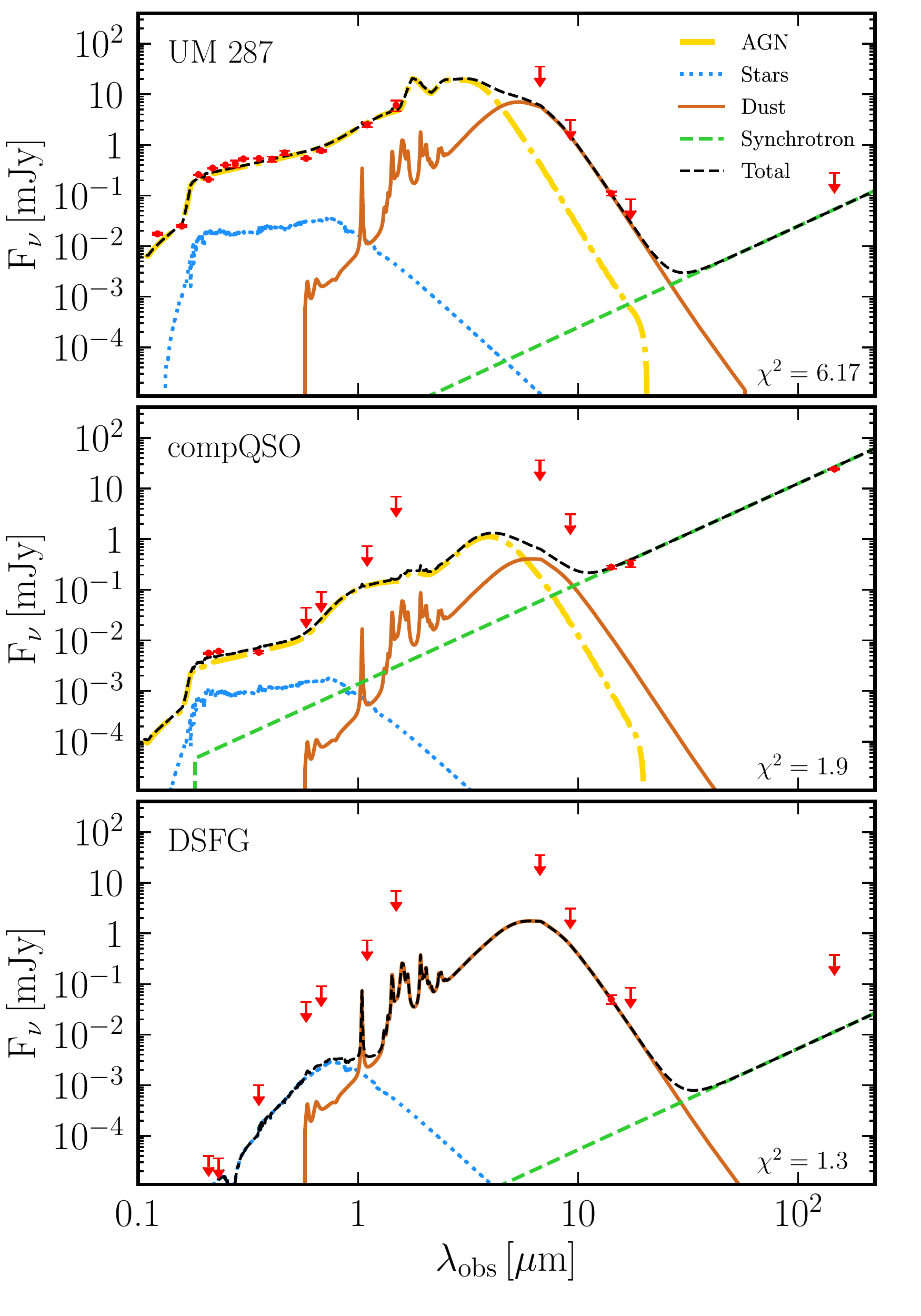}
		\caption{SED model fitting of all three sources. The wavebands considered are marked with red symbols, in which the 
		upper limits (\autoref{tab:table2}) are shown as downward arrows and points with error bars for significant measurements. The curves are the best-fit models determined by {\sc cigale}, color coded based on their constituent components, and the black dashed curves represent the sum of all components at a given wavelength. The reduced $\chi^2$ values of the best-fit models are given at the bottom-right corners. 
		}
		\label{fig:fig2}
	\end{center}
\end{figure}

Since our sources are infrared luminous with energy contributions from a mixture of black hole accretion and star formation, we utilize the SED fitting code {\sc cigale} \citep{Boquien:2019aa}, which is a self-consistent $\chi^2$ minimization modeling code that employs an energy-balance approach to account for absorption and re-radiation by dust grains both around AGN and within the interstellar medium. In short, the SEDs are modeled with four main components; 1) an AGN component parameterized as an accretion disk plus hot dust emission \citep{Fritz:2006aa}, 2) a stellar component from the host galaxy modeled under the assumption of an exponentially declining star formation history, a Chabrier initial mass function (\citealt{Chabrier:2003aa}), a modified starburst attenuation law \citep{Calzetti:2000aa, Leitherer:2002aa}, and a library of single stellar populations generated by \citet{Bruzual:2003aa}, 3) a dust emission component based on the models of \citet{Draine:2007ab,Draine:2014aa}, and 4) a synchrotron radiation component in radio estimated from the assumed correlation between infrared and radio. The rest-frame UV photometry is corrected for the foreground attenuation from the intergalactic medium following \citet{Meiksin:2006aa}. For each component the code offers a range of free parameters for the fitting and we adopt the setting used in Paper I \citep{Arrigoni-Battaia:2021aa}. \autoref{fig:fig2} shows the best-fit models with their associated reduced $\chi^2$. The relevant output physical properties based on these fittings are provided in \autoref{tab:table3}, {\ where we show that most properties are properly constrained except for the stellar mass of the Slug-DSFG, which is a result of mostly upper limits on its multi-wavelengths photometry.}

To understand the environmental impact to the source properties we also  estimate the halo properties. Following Paper I we estimate the halo masses using two approaches; First, we interpolate the halo mass $M_{\rm DM}$ - stellar mass $M_{\rm star}$ relations provided by \citet{Moster:2018aa} based on the redshifts of the three sources and adopt each of their stellar masses deduced from the {\sc cigale} fitting. The results are given in \autoref{tab:table3}, and the total is summed to 6.4$^{\ +27}_{-3.6}\times10^{12}$\,M$_\odot$. While the sum indeed suggests a massive halo of $10^{12}-10^{13}$\,M$_\odot$, this approach is clearly suffered from the uncertainties of the stellar mass estimates so we also adopt the second, dynamical approach. 

We estimate the dynamical mass of the system following the formalism of \citet{Tempel:2014aa}, which assumes a NFW dark matter density profile \citep{Navarro:1997aa} and a virialized, dynamically symmetric system,  allowing the link between the total velocity dispersion to the one-dimensional dispersion with a factor of $\sqrt{3}$. To compute the dark matter halo mass from the estimated dynamical mass we assume a cosmic baryon fraction obtained from {\it Planck}. If we only consider the three sources detected by ALMA, and again assuming the companion quasar being at the same redshift as that of UM287, we obtain a halo mass of $4.2\pm0.1\times10^{12}$\,M$_\odot$. If we also include other sources (compact sources `C' and `D' presented in \citet{Leibler:2018aa}), we obtain a halo mass of $1.3\pm0.4\times10^{13}$\,M$_\odot$. While very different in principle, the two approaches appear to converge on the fact that the system is likely to be hosted by a massive halo, with a total halo mass of $10^{12}-10^{13}$\,M$_\odot$, consistent with the clustering measurements of quasars at similar redshifts \citep{Rodriguez-Torres:2017aa}. Given the halo mass estimates we can calculate virial radius by simply following $R_{200}$ = $(3M_{200}/4\pi200\rho_c(z))^{1/3}$ where $\rho_c(z)$ is the critical density at the redshift of UM287 and $M_{200}$ is the halo mass estimated above. As a result we arrive at a virial radius of $R_{200}$ of about {\ 150-220\,kpc given the range of the halo mass}. Theoretically, the estimated halo mass range suggests that the Slug ELAN system is expected to live in a hot halo filled with shocked gas but may be fed by streams of cold gas from the intergalactic medium \citep{Dekel:2006aa}.

%FIGURE 2%

\begin{table*}[ht!]
    \begin{center}
        \caption{Physical properties obtained from the SED fitting}
        \label{tab:table3} 
        \begin{tabular}{lrrrll}
            \hline
            Quantity & UM287 & compQSO & Slug-DSFG & Units & Method\\
            \hline
            $L_{\text{IR}}$ & 5.9(0.3)$\times10^{12}$ & 3.0(0.2)$\times10^{11}$ & 1.3(0.5)$\times10^{12}$ & L$_\odot$ & integral of the dust SED 8-1000 micron\\
            $L_{\text{IR}}$$_{\text{tot}}$ & 6.0(0.4)$\times10^{13}$ & 1.9(0.3)$\times10^{12}$ & 1.3(0.5)$\times10^{12}$ & L$_\odot$ & integral of the total SED 8-1000 micron\\
            $L_{\text{bol}}$ & 6.8(0.5)$\times10^{13}$ & 1.8(0.1)$\times10^{12}$ & 5.0(2.4)$\times10^{11}$ & L$_\odot$ & integral of the whole SED\\
            $M_{\text{dust}}$ & 7.1(1.3)$\times10^{8}$ & 3.6(2.1)$\times10^{7}$ & 3.4(0.4)$\times10^{8}$ & M$_\odot$ & computed by CIGALE \citep{Draine:2014aa}\\
            $M_{\rm molgas}$ & 1.6(0.2)$\times10^{10}$ & $<4.3\times10^{9}$ & 1.2(0.4)$\times10^{10}$ & M$_\odot$ & estimated from the measured CO luminosities \\ 
            $\delta_{\rm GDR}$ & 23(5) & $<$118 & 34(12) & & Gas-to-dust ratios \\
            $M_{\text{star}}$ & 7.9(5.7)$\times10^{10}$ & 3.9(2.6)$\times10^{9}$ & 7.9(12.0)$\times10^{10}$ & M$_\odot$ & computed by CIGALE (exponential SFH)\\
            SFR & 399(70) & 19(3) & 182(108) & M$_\odot$ yr$^{-1}$ & computed by CIGALE (exponential SFH)\\
            M$_{\text{DM}}$ & 3.0($^{+7.0}_{-2.0}$)$\times10^{12}$ & 3.7($^{+0.9}_{-1.4}$)$\times10^{11}$ & 3.0($^{+26}_{-3}$)$\times10^{12}$ & M$_\odot$ & using $M_{\text{star}}$-$M_{\text{DM}}$ in \citet{Moster:2018aa}\\
            \hline
        \end{tabular}
        \begin{tabular}{l}
            The uncertainties are quoted in parentheses. The SFRs are average values over 100\,Myr.
    	\end{tabular}    
    \end{center}
\end{table*}

\section{Discussion} \label{sec:discussion}
\subsection{CO and {[CI]} line luminosity}\label{sec:lines}
To investigate the impact of dense environment on ISM-rich galaxies detected by our ALMA observations, we first discuss relations between line properties and infrared luminosity, and put them into the context with respect to other galaxy populations. To do so, we have compiled a list of other galaxy samples that have had reported measurements similar to those of our sources, meaning CO(4-3), [CI](1-0), line width in FWHM, and infrared luminosity. This list includes $z=2-5$ submillimeter galaxies \citep{Bothwell:2013lp,Alaghband-Zadeh:2013aa,Bothwell:2017aa,Birkin:2021aa}, quasars \citep{Carilli:2013aa,Bischetti:2021aa}, and one star-forming galaxy on the main sequence \citep{Brisbin:2019aa}, $z\sim1$ star-forming galaxies on the main sequence \citep{Valentino:2018aa,Bourne:2019aa,Boogaard:2020aa}, and local galaxy samples with {\it Herschel} FTS measurements \citep{Kamenetzky:2016aa}. 

Since the compiled list consists of different studies focusing on various types of galaxies using at times different analysis methodologies to estimate source properties, such as infrared luminosity, it requires certain efforts to try homogenizing the measurements as much as possible. First, to avoid inconsistency due to different assumptions of cosmology, in all cases we deduce line luminosity based on the reported line flux densities, their associated uncertainties and the assumed cosmology in this paper. The line luminosity in K\,km\,s$^{-1}$\,pc$^2$ is derived using the standard equation $L^\prime_{\rm CO}=3.25\times10^7 S_{\rm CO}\Delta v\nu_{\rm obs}^{-2}D_{\rm L}^2(1+z)^{-3}$ \citep{Solomon:2005aa}, where $S_{\rm CO}\Delta v$ is the total line flux in Jy\,km\,s$^{-1}$, $\nu_{\rm obs}$ is the observed line frequency in GHz, and $D_{\rm L}$ is the cosmological luminosity distance in Mpc. Secondly, we convert all the reported infrared luminosity to the one defined as the total luminosity at 8--1000\,$\mu$m, for which we assume $L_{\rm FIR(40-400\mu m)}$ = $0.7\times L_{\rm IR(8-1000\mu m)}$ and $L_{\rm IR(8-1000\mu m)}$ = $1.9\times L_{\rm FIR(42-122\mu m)}$ \citep{Carilli:2013aa}. 

\begin{figure*}[ht!]
	\begin{center}
		%\leavevmode
		\includegraphics[scale=0.88]{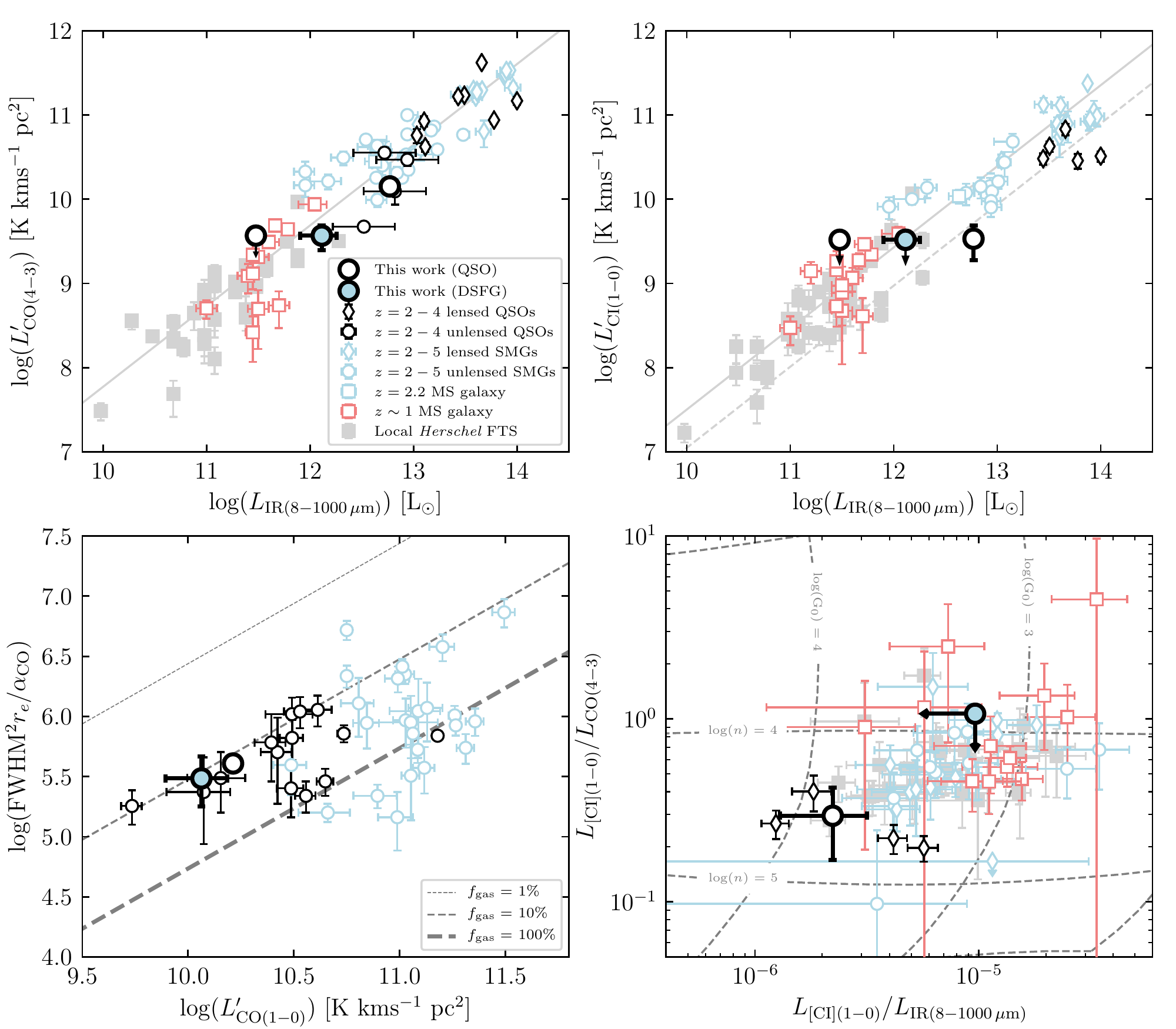}
		\caption{{\it Top}: Comparisons between line luminosity and infrared luminosity of the sources detected by the ALMA observations toward the Slug ELAN, as well as a compiled list of galaxy samples in the literature in both the nearby and high-redshift universe (See \autoref{sec:lines} for the included samples). The solid lines are best fit linear models to the star-forming galaxies, and the dashed one in the top-right panel shows the fit with the slope fixed but to the quasars only, which is a factor of 3$\pm$1 lower. {\it Bottom-left}: A diagram to effectively visualize the gas fraction in sample galaxies via a dynamical mass method (\autoref{sec:gf}), indicating that on average quasars, including UM287 measured in this work, have a lower gas mass fraction compared to SMGs at similar redshifts. The Slug-DSFG appears to be at the lower end of the gas mass fraction, consistent with the top-left panel showing slightly under-luminous nature for the Slug-DSFG relative to the SMGs, which may be due to environmental effects. {\it Bottom-right}: A luminosity ratio diagram that helps visualize differences in physical parameters of the PDR in different galaxy populations. The model curves are based on the {\sc pdr toolbox} described in \autoref{sec:pdr}. We find a denser field with stronger ultra-violet radiations in the PDR of quasars including UM287. 
		}
		\label{fig:fig3}
	\end{center}
\end{figure*}

Comparisons of various measurements based on the stated calculations are now plotted in \autoref{fig:fig3}. In the following we discuss them in detail. Note that the samples of $z=2-5$ galaxies include both lensed and unlensed sources, which are clearly separated in the plots. Since the lensing magnifications of all the reported lensed galaxies are assumed to be the same for both emission lines and infrared continuum, and as shown later the linear correlations have slopes consistent with unity, for clarity we plot the observed values without correcting for lensing.

%FIGURE 3%

In the top panels of \autoref{fig:fig3} we plot comparisons between line luminosity and infrared luminosity, which provide insights into the conditions of interstellar medium and star formation in these galaxies. In both CO(4-3) and [CI](1-0), when considering all data points, we find a linear correlation in log-log scale with a best-fit slope of unity and a scatter of 0.3\,dex. Upon closer inspections of each correlation, we find that for CO(4-3), both quasars and DSFGs lie on the general correlation. While for [CI](1-0), on the other hand, there is evidence of quasars being under luminous at fixed infrared luminosity. By fixing the slope we find the best-fit linear model for quasars is a factor of 3$\pm$1 lower in normalization than that of dusty star-forming galaxies, possibly suggesting either a more efficient star formation or a reduced amount of molecular gas in quasars, or both. Although as can be seen the total number of [CI](1-0) measurements on quasars is relatively small.

The different relationships between quasars and DSFGs seen in CO(4-3) and [CI](1-0) can be understood in their differences in excitation conditions of the ISM {\ (e.g., \citealt{Banerji:2017aa,Fogasy:2020aa})}. {\ That is, if both quasars and DSFGs lie on the same $L^\prime_{\rm CO(4-3)}$-$L_{\rm IR}$ correlation to start with, the different $L^\prime_{\rm CO(4-3)}/L^\prime_{\rm CO(1-0)}$ luminosity ratios (0.46 and 0.87 for DSFGs\footnote{We assume the ratio for the SMGs is applicable to the $z\sim1$ main-sequence star-forming galaxies in our compiled list.} and quasars respectively; \citealt{Carilli:2013aa}) would mean that the $L^\prime_{\rm CO(1-0)}$-$L_{\rm IR}$ linear relation for quasars would be a factor of about 2 lower than that for star-forming galaxies.} If we instead adopt the latest conversion from SMGs for star-forming galaxies, which is 0.32 \citep{Birkin:2021aa}, the difference would grow to a factor of about 3, even closer to our results since both CO(1-0) and [CI](1-0) line luminosity predominantly reflects molecular gas masses. Our results are consistent with recent report by \citet{Bischetti:2021aa}, who found CO(1-0) emission lines being systematically under luminous by a factor of $\sim4$ compared to main-sequence galaxies and SMGs for a sample of $z=2-5$ WISE-SDSS selected hyper-luminous quasars.

One possible concern over the enhanced infrared luminosity at fixed line luminosity for quasars is the effectiveness of removing the AGN contribution. While the AGN contributions to the infrared luminosity have been carefully removed for our sources via the modelling of {\sc cigale}, those for other quasars in the compiled list have not \citep{Carilli:2013aa,Bischetti:2021aa}, at least not specifically, although various methods were adopted to account for this effect. For example \citet{Carilli:2013aa} used a dusty star-forming galaxy SED template of Arp220 to estimate the infrared luminosity by fitting it to the submillimeter photometry, thus in principle the quoted infrared luminosity do not contain much AGN contributions. Furthermore, recent studies exploiting {\it Herschel} photometry on these lensed quasars have argued for star formation being the dominant contributor to the infrared luminosity, based on them having consistent infrared-to-radio correlation with that of star-forming galaxies \citep{Stacey:2018bb}. In addition, studies with careful SED modelling to remove the AGN contributions on a sample of $z>1$ quasars have found similar results of under luminous CO \citep{Perna:2018aa}. All these suggest that our results are not sensitive to the exact techniques used to remove the AGN contributions to the infrared luminosity.

For the sources detected in the Slug ELAN, both UM287 and the Slug-DSFG appear to lie close to the lower bound of the scatter based on the general correlations, which could suggest environmental effects on their cold gas reservoir. We look at this in more detail and continue this part of discussion in the next section.

\subsection{Molecular gas fraction and star formation efficiency}\label{sec:gf}

The results of under luminous CO and [CI] suggest higher efficiency of star formation, which could be due to relatively lower gas masses or relatively higher {\ star-formation rates (SFRs)} compared to the general SMG and main-sequence galaxies, or both. We investigate these possibilities by first looking at the gas mass fraction. We adopt both the dynamical and the direct methods; For the dynamical method we compare CO line widths and luminosity, which can be linked via a dynamical mass estimate expressed as $M_{\rm dyn} = C(v_c/{\rm sin}(i))^2\times R/G$, under the assumption of a rotation dominated system, where $G$ is the gravitational constant, $v_c$ is the circular velocity, $R$ is the representative radius, $C$ is the correction factor depending on the mass distributions and the adopted representative radius, and $i$ is the inclination angle. Assuming an exponential disk profile with a scale length $h$ and a half-light radius $r_e$, it has been shown that $v_c = {\rm FWHM_{\rm CO}}/2$ and $R=3h=1.79r_e$\footnote{$r_e=1.68h$ for an exponential disk profile \citep{Chen:2015aa}} are good approximations for spatially unresolved CO observations \citep{deBlok:2014aa}. Under this assumption of an exponential disk $C$ is about 2 \citep{Binney:2008aa}. In addition, dynamical masses can be linked to molecular gas masses with a gas mass fraction defined as $f_{\rm gas} = M_{\rm gas}/M_\ast$ where $M_\ast$ is stellar mass, and a dark matter fraction $f_{\rm DM}$, such that $M_{\rm dyn}=(M_\ast+M_{\rm gas})/(1-f_{\rm DM})=M_{\rm gas}(1+f_{\rm gas})/(f_{\rm gas}(1-f_{\rm DM}))$. Assuming gas mass is dominated by molecular gas then gas masses are simply $M_{\rm gas} = M_{\rm molgas} = \alpha_{\rm CO}L^\prime_{\rm CO(1-0)}$ where $\alpha_{\rm CO}$ is effectively a mass to light ratio to molecular hydrogen mass. With all these we can then re-write the equation of dynamical mass estimates to 

\begin{equation}\label{eqn:eqn1}
    \begin{split}
    & {\rm log}(\frac{{\rm FWHM_{\rm CO}}^2r_e}{\alpha_{\rm CO}}) = \\
    & {\rm log}(L^\prime_{\rm CO(1-0)}) - {\rm log}(\frac{2.3\times10^5f_{\rm gas}(1-f_{\rm DM})\times1.79\times2}{4{\rm sin}^2(i)\times(1+f_{\rm gas})}) = \\
    & {\rm log}(L^\prime_{\rm CO(1-0)}) - {\rm log}(\frac{3.1\times10^5f_{\rm gas}(1-f_{\rm DM})}{(1+f_{\rm gas})})
    \end{split}
\end{equation}
, where FWHM is in km s$^{-1}$ and $r_e$ in kpc, and we adopt $\langle{\rm sin}^2(i)\rangle=2/3$ \citep{Tacconi:2008p9334}. For dark matter fraction, we adopt $f_{\rm DM}=0.12$ based on the latest dynamical studies of $z\sim2$ star-forming galaxies with analyses of high-quality rotational curves \citep{Genzel:2020aa}. In the bottom-left panel of \autoref{fig:fig3} we plot three lines representing three gas mass fractions respectively, in which we also plot our measurements along with those from the $z=2-5$ unlensed sources in the literature. For this exercise we also include the CO(3-2) measurements on a sample of $z=2.5-3.0$ quasars from \citet{Hill:2019aa}. To convert $L^\prime_{\rm CO(4-3)}$ to $L^\prime_{\rm CO(1-0)}$ we adopt the most up-to-date luminosity ratios reported in the literature, meaning $r_{41}=L^\prime_{\rm CO(4-3)}$/$L^\prime_{\rm CO(1-0)}$ ratios of 0.87 and 0.32 for quasars and SMGs, respectively \citep{Carilli:2013aa,Birkin:2021aa}. For the CO(3-2) measurements on quasars reported in \citet{Hill:2019aa} we adopt a $r_{31}=0.97$ again based on \citet{Carilli:2013aa}.
In addition, we adopt $\alpha_{\rm CO}=1.0${\ \,M$_\odot$(K\,km\,s$^{-1}$\,pc$^2$)$^{-1}$} and $r_e=3$\,kpc for SMGs and quasars \citep{Chen:2017aa,Tuan-Anh:2017aa,Calistro-Rivera:2018aa,Badole:2020aa,Bischetti:2021aa}. Note these adopted values are averages, and each of which comes with a range of uncertainties that are beyond the scope of this paper. The point of this exercise is to look for possible systematic differences between different galaxy populations thus the averaged values are sufficient. As we can see, in general the dusty star-forming galaxies have a higher gas fraction compared to that of quasars at similar redshifts. By fitting \autoref{eqn:eqn1} to the data we deduce a gas fraction of dusty star-forming galaxies of 0.5$\pm$0.1, compared to 0.2$\pm$0.1 for quasars, although each has about 0.3\,dex scatter. For UM287 we find a gas fraction of 0.2$\pm$0.1, consistent with the average of the quasar sample, and 0.1$\pm$0.1 for the Slug-DSFG, on the lower end of the SMG sample.

We now look at the direct method of computing gas mass fractions by taking the fractions between the molecular gas masses and stellar masses. By adopting the corresponding luminosity ratios, $\alpha_{\rm CO}$, and the $L^\prime_{\rm CO(4-3)}$ given in \autoref{tab:table1}, we estimate molecular gas masses of $M_{\rm gas}=1.6\pm0.2\times10^{10}M_\odot$ and $1.2\pm0.4\times10^{10}M_\odot$ for UM287 and the Slug-DSFG, respectively, corresponding to gas fractions of $0.2\pm0.2$ and $0.1\pm22$. We also estimate the weighted averaged gas fraction for the SMGs to be $0.5\pm0.1$, in excellent agreement with that based on the dynamical method and the reported values for SMGs in the literature \citep{Bothwell:2013lp,Birkin:2021aa}. The large uncertainties for both UM287 and Slug-DSFG are due to the fact that the stellar masses are not well constrained, in particular for the Slug-DSFG. Despite that, their general behavior appears in agreement with the dynamical method, meaning that the gas fractions estimated from line emission tend to be lower for sources in the Slug ELAN, compared to other dusty star-forming galaxies at similar redshifts, suggesting that a lower gas mass could be the reason for them being under luminous as discussed in \autoref{sec:lines}. On the other hand, given their SFRs and stellar masses UM287 and Slug-DSFG are located about 1-2 times the scatter above the main sequence so their SFRs are slightly enhanced. That is, the reasons of them being under luminous in line luminosity could be due to both smaller amount of gas reservoir and higher SFRs compared to star-forming galaxies at similar redshifts.

Adopting the SFRs given in \autoref{tab:table3}, which are averaged over 100\,Myr based on the {\sc cigale} SED models, the gas depletion timescales, defined as $M_{\rm gas}$/SFR, are 41$\pm$8\,Myr and 64$\pm$43\,Myr for UM287 and Slug-DSFG, respectively. The timescales would be 19$\pm$10\,Myr and 84$\pm$95\,Myr if we instead adopting the IR-base SFRs with the \citet{Kennicutt:2012aa} conversion. The typical depletion timescale for SMGs is about 200-300\,Myr \citep{Dudzeviciute:2020aa,Birkin:2021aa}, which puts the Slug-DSFG at the lower end of the depletion time, again suggesting that it has relatively less gas which could be due to its extreme environments.

It is worth noting that we can also estimate molecular gas masses using the measured {[CI]} luminosity. By following equation 6 of \citet{Bothwell:2017aa} with a correction of helium contribution of 1.36 we find a [CI]-base molecular gas mass of $3.4\pm1.4\times10^{10}$M$_\odot$ and $<3.2\times10^{10}$M$_\odot$ for UM287 and the Slug-DSFG, respectively; these are approximately {\ a} factor of two larger than the CO-base estimates. If the [CI]-base molecular gas masses are closer to the truth, then we will have to double the gas fraction and the depletion time, putting UM287 closer to the star-forming galaxies at similar redshifts under this context. However, like the known issue of the $\alpha_{\rm CO}$ conversion factor, the {[CI]}-base estimates are also prone to the unknown nature of the [CI]/H2 abundance ratio, subject to variations of more than a factor of two. A standard ratio of $3\times10^{-5}$ was adopted for our calculation, which was estimated based on observations of M82 \citep{Weis:2003aa}. However, this ratio is known theoretically to vary by orders of magnitude under different local physical conditions such as density and radiation field (e.g., \citealt{Papadopoulos:2014aa}), and observationally for high-redshift galaxies it lacks constraints that are free from other key assumptions such as $\alpha_{\rm CO}$ and gas-to-dust ratio which are shown to heavily influence the results \citep{Valentino:2018aa}. Given the circumstances it appears more appropriate to compare physical quantifies that are derived from similar measurements. Since most molecular gas masses reported in the literature and used for comparisons in this work are based on CO measurements, we opt to adopt the CO-base estimates.

In summary, we tend to interpret that, like other $z>1$ quasars or AGN \citep{Perna:2018aa,Man:2019aa,Bischetti:2021aa}, the gas fraction of UM287 is low and it is using up its gas reservoir very soon in less than 50\,Myrs. On the other hand the Slug-DSFG appears to have less gas compared to typical SMGs at similar redshifts, perhaps due to the extreme environments around the ELAN that facilitate some level of gas stripping or strangulation. We extend this part of discussion in a later section.

\subsection{Physical parameters of the PDR}\label{sec:pdr}
Line detection of multiple transitions and species in far-infrared and submillimeter allows simple modelling of physical conditions of the photo-dissociation region (PDR), in particular for high-redshift galaxies the luminosity-weighted mean hydrogen nucleus number density, $n$, and the incident far-ultraviolet intensity, $G_{\rm 0}$. 

In the bottom-right panel of \autoref{fig:fig3} we plot the luminosity ratios between $L_{\rm [CI](1-0)}$/$L_{\rm CO(4-3)}$ and $L_{\rm [CI](1-0)}$/$L_{\rm IR}$. We observe that UM287 is located at a similar {\ locus} as that of other quasars at similar redshifts, however it deviates notably from the main {\ locus} where most of the star-forming galaxies are located, such that the quasars have smaller values in both ratios. The upper limits of the Slug-DSFG put it in between the star-forming galaxies, and the $z\sim1$ main-sequence galaxies appear to offset from the SMGs toward higher $L_{\rm [CI](1-0)}$/$L_{\rm IR}$ ratios.

To model these ratios we adopt the {\sc pdr toolbox} \citep{Kaufman:2006aa,Pound:2008aa}, which is a grid-base , minimizing $\chi^2$ model space fitting tool built upon various theoretical studies \citep{Tielens:1985aa,Kaufman:1999aa}. The expected ratios given a range of constant $n$ and $G_0$ are plotted in \autoref{fig:fig3}. We find that the majority of the star-forming galaxies have log($n$)=4-4.5\,cm$^{-3}$ and log($G_0$)=2.5-3.5\,Habing, and the $z\sim1$ main-sequence galaxies have a similar range of $n$ but they are slightly lower in $G_0$. The quasars including UM287 have on average about log($n$)=4.5\,cm$^{-3}$ and log($G_0$)=3.5\,Habing, at the higher end of both $n$ and $G_0$, suggesting denser and stronger far-ultraviolet radiation field.

Our results on high-redshift star-forming galaxies are consistent with those reported recently by \citet{Valentino:2020aa}, who have studied similar line ratios on a sample of high-redshift galaxies in which they have also found a similar range of $n$ and $G_0$ in both star-forming galaxies and AGN or quasar samples. Our measurements on UM287 add to the existing measurements supporting that the PDR in quasars is denser and stronger in far-ultraviolet radiation field compared to star-forming galaxies at similar redshifts.

We however caution that this exercise assumes that all line and dust emission originates from the PDR. The contributions from the X-ray dominated regions (XDR) on different transitions of CO lines for high-redshift quasars remain unclear. While some studies have found low- and mid-$J$ ($J\le7$) CO to be predominantly emitted from the PDR \citep{Pensabene:2021aa}, other studies have shown XDR could make significant contributions for CO transitions as low as $J=5$ \citep{Uzgil:2016aa,Bischetti:2021aa}. It is however unclear for CO(4-3). If a significant fraction of CO(4-3) is powered by the XDR, this would increase the [CI](1-0)-to-CO(4-3) ratios thus lower $n$, moving it closer to the conditions of star-forming galaxies.

\subsection{Gas-to-dust ratios}\label{sec:gdr}
With the estimates of molecular gas mass presented in \autoref{sec:gf} and the dust masses derived from {\sc cigale} (\autoref{subsec:sed}), we can now look at the gas-to-dust ratios $\delta_{\rm GDR}$. As shown in \autoref{tab:table3}, we find the two sources that have significant measurements in both quantities, UM287 and Slug-DSFG, have their $\delta_{\rm GDR}$ about 20-30. This is significantly lower than the typical values of $\sim$100 estimated for other samples of quasars and dusty galaxies at similar redshifts (e.g., \citealt{Shapley:2020aa,Birkin:2021aa,Bischetti:2021aa}), but consistent with what has been found for the quasars and AGN in another ELAN \citep{Arrigoni-Battaia:2021aa}.

As a check we have also fit the 2\,mm photometry with simpler modified black body (mBB) models. This is justified such that given the redshift of these sources the 2\,mm photometry probes the Rayleigh-Jeans tail which is the part mainly determined by dust masses. Under the typical assumptions such that emission is isotropic over a spherical surface, optically thin, and single temperature, the model can be described as 
\begin{equation}
    S_{\nu_o} \propto \nu_e^\beta B(\nu_e,T_d)
\end{equation}
where $S_{\nu_o}$ is the flux density at the observed frequency, $\nu_e$ is the rest frequency such that $\nu_e$ = (1+$z$)$\nu_o$, $\beta$ is dust emissivity spectral index, and $B(\nu_e,T_d)$ the rest-frame Planck function with a temperature $T_{\rm d}$. We can then deduce dust masses using 
\begin{equation}
    M_{\rm dust} = \frac{D_L^2S_{\nu_o}}{(1+z)\kappa_{\nu_e}B(\nu_e,T_d)} 
\end{equation}
where $D_L$ is the luminosity distance at the source redshift $z$ and $\kappa_{\nu_e}$ is the rest-frame dust mass absorption coefficient. In this case we adopt $\kappa_{\nu_e}=0.431$\,cm$^2$g$^{-1}$ at $\nu_e=353$\,GHz (850\,$\mu$m) based on the dust model of Li \& Draine (2001), for which a similar model was adopted in the {\sc cigale} fitting. Assuming a range of dust emissivity spectral index of $\beta = 1.5-2.0$ and dust temperature of $T_{\rm d}=20-50$ that are appropriate for SMGs and quasars \citep{Beelen:2006aa, Carilli:2013aa, Dudzeviciute:2020aa}, we find dust masses of 2-10$\times10^{8}$ and 1-5$\times10^{8}$ M$_\odot$ for UM287 and Slug-DSFG, in good agreement with the estimates from {\sc cigale}.

%FIGURE 4%
\begin{figure}[ht!]
	\begin{center}
		%\leavevmode
		\includegraphics[scale=0.75]{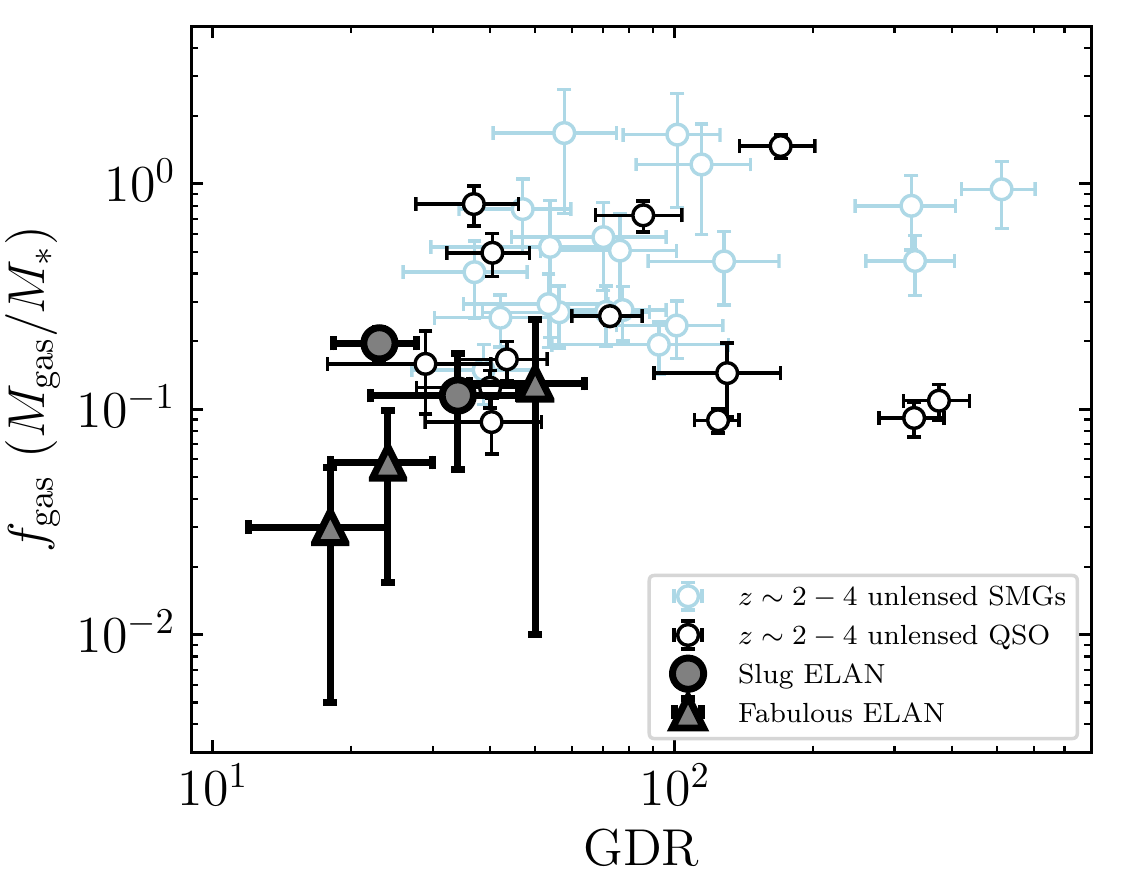}
		\caption{A plot contrasting gas-to-dust ratios (GDR) and gas fractions, where similar to \autoref{fig:fig3} the blue symbols represent the $z\sim2-4$ unlensed SMGs, the empty black for $z\sim2-4$ unlensed QSOs, and the filled black for sources in ELAN, {\ including the two sources in the Slug ELAN (circles) and the three QSO and AGN in the Fabulous ELAN (triangles; \citealt{Arrigoni-Battaia:2021aa})}.} 
		\label{fig:fig4}
	\end{center}
\end{figure}

With now two independent methods converging on dust mass estimates, it appears that the sources located within ELAN have both their gas fractions and gas-to-dust ratios on average lower than the typical SMGs found at similar redshifts. This intriguing finding motivated a detailed comparison between our sources in the Slug ELAN and other unlensed SMGs \citep{Birkin:2021aa} and unlensed quasars \citep{Hill:2019aa,Bischetti:2021aa} at similar redshifts ($z\sim2-4$) reported in the literature. To increase the sample size of sources in ELAN, in this exercise we also include three quasars and AGN found in the $z=3.16$ Fabulous ELAN \citep{Arrigoni-Battaia:2021aa}, where CO(5-4) measurements were obtained and the dust masses were estimated using the same SED fitting parameters for {\sc cigale}.

Since both gas fraction and gas-to-dust ratios may be affected significantly by a few underlying assumptions, it is important to homogenize the methodology if possible, or to understand the possible systematic differences if different methods were used. For gas-to-dust ratios, given the fact that the dynamical method can reproduce well the averaged fractions of the direct method (\autoref{sec:gf}), we adopt the dynamical method for all the sources considered in this exercise, where the line widths and line intensities are provided. In practice, we fit the measurements of all the sources using \autoref{eqn:eqn1}, under the same assumptions of parameters as those described in \autoref{sec:gf}. This procedure ensures that the differences seen in gas fractions are not caused by the different assumptions of $\alpha_{\rm CO}$ and sizes. Note that for the three sources that are found in the Fabulous ELAN, the spatially resolved CO(5-4) measurements in fact showed that they have half-light radius of 3-5\,kpc, which are larger than our assumed 3\,kpc for the dynamical method, and would further lower the gas fractions of these sources. Since we are interested in the averaged values we opt to adopt the same sizes.

On the other hand, for gas-to-dust ratios we adopt the reported CO line flux densities and compute gas masses based on the assumed cosmology of this paper and the adopted luminosity ratios presented in \autoref{sec:gf}. For dust masses we simply adopt the reported values in the literature for the SMGs and quasars \citep{Birkin:2021aa,Hill:2019aa,Bischetti:2021aa}. However, we caution that each study employed different methods for the infrared SED fitting and deducing dust masses, such as modified black body models or another SED fitting code {\sc magphys} \citep{da-Cunha:2008aa}, adopting a range of dust emissivity spectral index ($\beta=1.6-2.0$) and dust mass absorption coefficient ($\kappa_{\nu_e}=0.431-0.77$\,cm$^2$g$^{-1}$ at 850\,$\mu$m). While a detailed comparison of different dust modeling on these sources is outside the scope of this paper, previous studies using other $z\sim1-2$ galaxy samples have shown these different models to impact the results on a level of about a factor of two (e.g., \citealt{Magdis:2012aa}). 

%FIGURE XX%
\begin{figure*}%[h!]
	\begin{center}
		%\leavevmode
		\includegraphics[width=1.0\textwidth]{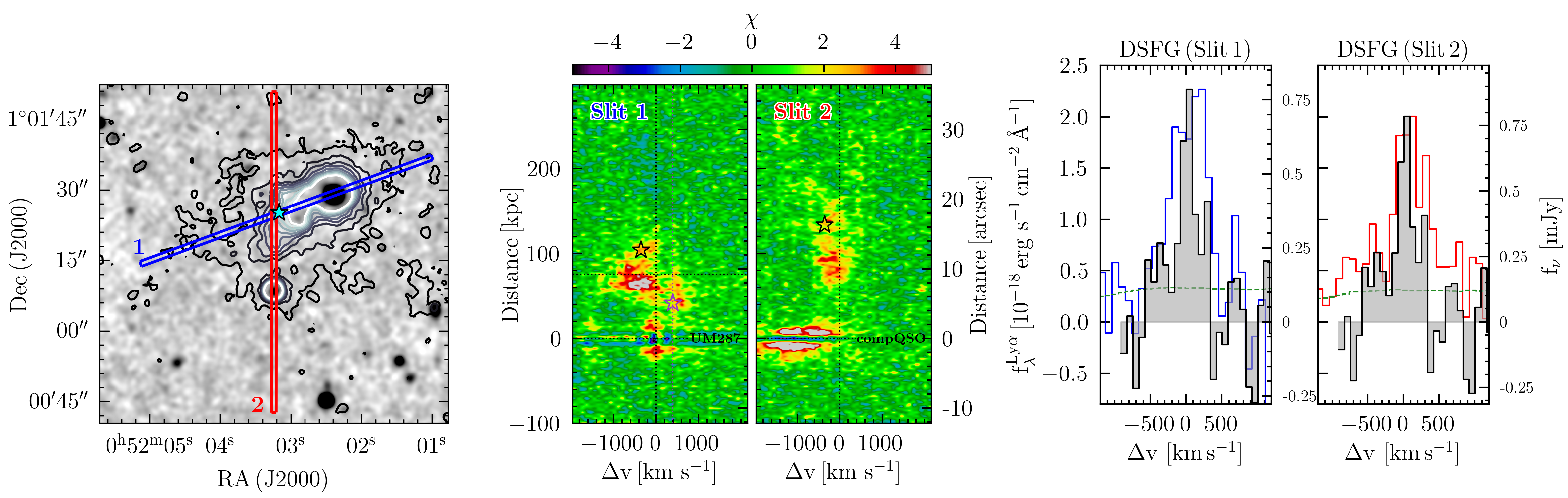}
		\caption{The spatial and spectral location of the DSFG in comparison to the Slug ELAN. \emph{Left:} HAWK-I/$J$-band map with overimposed the two Keck/LRIS slits of $1\arcsec$ (slit 1 in blue and slit 2 in red; \citealt{Arrigoni-Battaia:2015aa}) and the contours for the Ly$\alpha$ emission in the range S/N~$=2-47$ in steps of 5 (\citealt{Cantalupo:2014aa}). The discovered DSFG (star) is inside slit 1 and right next slit 2. \emph{Middle:} Keck/LRIS 2D $\chi$ spectra at the location of the Ly$\alpha$ line (\citealt{Arrigoni-Battaia:2015aa}). The reference velocity is the systemic redshift of the quasar UM287 ($z=2.2825$).The DSFG is indicated by the black star. For reference, the position of source ``c'' (purple star) and the average \ion{He}{2} velocity (vertical purple dashed line) are indicated (\citealt{Cantalupo:2019aa}). 
		\emph{Right:} Comparison of the 1D spectrum in Ly$\alpha$ (extracted within a $2\arcsec$ aperture) and CO(4-3) for the DSFG. The reference velocity is the systemic redshift of the DSFG ($z=2.2786$). {The two Ly$\alpha$ lines are plotted at the same flux density scale indicated on the left y-axis, while the CO(4-3) lines for comparison are normalized to the peak of each Ly$\alpha$ line and their flux density scales are shown on each of their right y-axis.}} 
		\label{fig:fig5}
	\end{center}
\end{figure*}

We plot the results in \autoref{fig:fig4}, where the sources in ELAN appear offset toward lower ends in both gas fraction and gas-to-dust ratio. Quantitatively, the medians with bootstrapped errors for gas fraction are $0.45\pm0.10$, $0.16\pm0.11$, and $0.12\pm0.04$, for SMGs, quasars, and sources in ELAN respectively, and $74\pm12$, $73\pm34$, and $24\pm7$ for gas-to-dust ratios, respectively. That is, given the uncertainties the differences appear significant ($>3$\,$\sigma$) between SMGs and the sources in ELAN on both gas fractions and gas-to-dust ratios, while the difference in gas fraction between SMGs and quasars appear marginal ($\sim$2\,$\sigma$).

As suggested by the analyses in \autoref{sec:gf}, the general trend of quasars and AGN having lower gas fractions than SMGs is consistent with the evolutionary scenario where the quasar phase follows the SMG phase such that the quasars are closer in using up their cold ISM reservoir \citep{Toft:2014aa}, which has also been suggested by other studies of molecular gas content in $z\sim2$ quasars \citep{Perna:2018aa,Hill:2019aa,Bischetti:2021aa}.

On the other hand, while possible systematic differences in the reported dust masses need to be understood further, the apparently lower gas-to-dust ratios between sources in ELAN and the comparison samples of SMGs and quasars is intriguing. Given the observed correlation between gas-to-dust ratios and gas-phase metallicities in both nearby and high-redshifts galaxies \citep{DeVis:2019aa, Shapley:2020aa}, it may be that the sources in ELAN are more metal rich. However the measured ratios would imply a metallicity that is about two times the solar metallicity based on correlations provided by \citet{Leroy:2011aa} and \citet{DeVis:2019aa}. 

Physically this may be a result of gas stripping caused by active dynamical interactions in dense environments, leading to a higher gas-phase metallicity which is essentially a oxygen-to-hydrogen abundance ratio. This scenario may find support from the recent findings of more compact sizes of dust continuum compared to gas and stellar emission in dusty galaxies (e.g.,  \citealt{Chen:2017aa,Calistro-Rivera:2018aa,Tadaki:2019aa}), where stronger stripping effects on gas than dust may be expected, which would lower the gas-to-dust ratios. On the other hand, it could also be due to strangulation where the gas supply is made inefficient by the hot halos, such that metallicity increases as a function of time with continuous star formation as well as decreasing the gas-to-dust ratios \citep{Hirashita:1999aa,Peng:2015aa}. Finally, it is known that the circumgalactic medium of $z\sim2-3$ quasars is already metal enriched (about one third of solar; \citealt{Lau:2016aa,Fossati:2021aa}), thus depending on the timescales it is also possible that the accretion of enriched gas onto galaxies helps elevate the gas-phase metalicity. We further extend this discussion in the later sections. 

\subsection{Ly$\alpha$-emitting gas from the Slug-DSFG}
To understand the origin of the Ly$\alpha$ nebula, in \citealt{Arrigoni-Battaia:2015aa}) we measured and modeled the properties of Ly$\alpha$, He~{\sc ii}, and C~{\sc iv} lines using two slits of deep LRIS spectra taken from the Keck telescope. Their respective locations and orientations are shown in \autoref{fig:fig5}. Coincidentally, the newly discovered Slug-DSFG, marked as stars in the figure, is located almost right at the intersection of the two slits. While one of the slits (slit 2) covers only part of the source, the other one (slit 1) goes through the source entirely. What is even more remarkable is that Ly$\alpha$ is clearly emitting from the locations of the Slug-DSFG, in both real and velocity spaces.

We first extract spectra using the same circular aperture with a size of 2$''$ in diameter as that used for extracting optical photometry (\autoref{subsec:sed}). In the right panel of \autoref{fig:fig5} we show the line profiles of Ly$\alpha$ from the two slits, and compare them to the CO(4-3) line normalized to the peaks of Ly$\alpha$. We fit the line profiles with single Gaussian models and find Ly$\alpha$ line luminosities of $8.5\pm1.1\times10^{41}$ and $9.8\pm1.2\times10^{41}$ erg s$^{-1}$ with widths in FWHM of 632$\pm$70 and 1200$\pm$120 km\,s$^{-1}$ from slit 1 and 2, respectively. Since slit 2 covers Slug-DSFG only partially, in principle the line luminosity should be smaller than that measured from slit 1, and the line widths should agree with each other. The fact that we are getting more emission and a wider line width from slit 2 could mean that the aperture size is too large such that it also covers emission coming from the gas surrounding the Slug-DSFG. {\ This could be due to additional gas motions such as inflow or outflow, and as discussed in the later sections we attribute it mainly to gas ram pressure stripping.} Indeed, if we extract the spectra using a 1$''$ circular aperture instead, the line width appears consistent with that obtained from slit 1, and the line luminosity from slit 2 is indeed smaller than that from slit 1. We therefore adopt the Ly$\alpha$ line luminosity and width from slit 1, which are $8.5\pm1.1\times10^{41}$\,erg\,s$^{-1}$ and 632$\pm$70\,km\,s$^{-1}$, and conclude that the broader component from slit 2 is due to the surrounding gas that is presumably undergoing gas mixing as part of the stripping process. The good agreement on the systemic velocity between Ly$\alpha$ and the line detected by our ALMA data strengthens the argument that the line is CO(4-3).

\begin{figure*}
	\begin{center}
		%\leavevmode
		\includegraphics[width=1\textwidth]{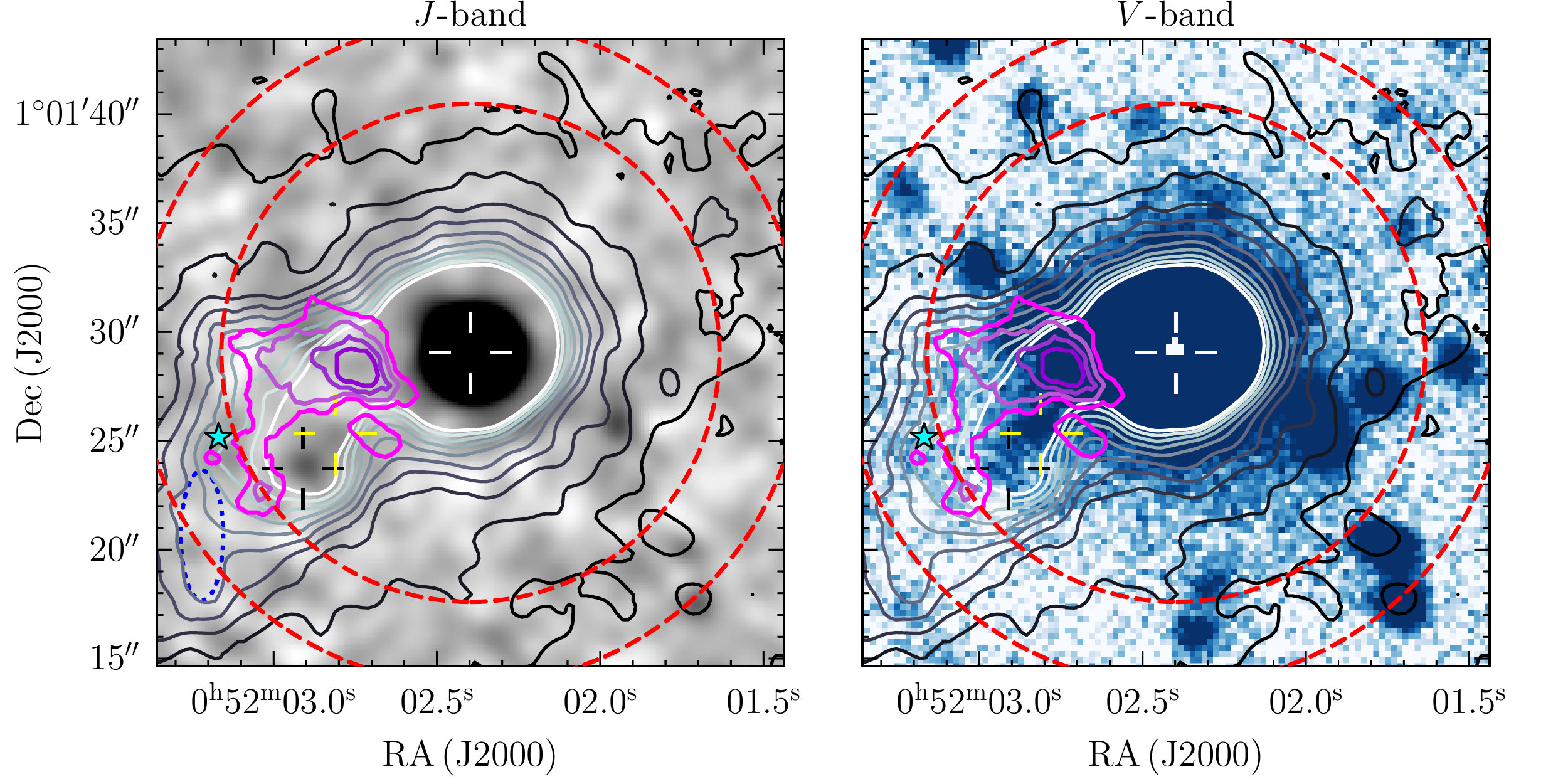}
		\caption{\emph{Left:} HAWKI $J$-band data with overimposed the Ly$\alpha$ contours as in Figure~\ref{fig:fig5} and the 2, 4, 6, 8 S/N contours for the \ion{He}{2} emission associated with source ``C'' (magenta to purple; \citealt{Cantalupo:2019aa}). The DSFG is indicated by the cyan star, while UM287 {\ and the two continuum sources co-locating with the bright tail (detail given in \autoref{sec:brightail})} are indicated with white, black, and yellow crosshairs, respectively. The dotted blue ellipse indicates the area used in the calculation of the stripped material (section~\ref{sec:strip}). \emph{Right:} Same as the left panel, but for the LRIS $V$-band. Continuum emission is evident within the brightest portion of the ELAN.}
		\label{fig:fig6}
	\end{center}
\end{figure*}

We investigate the origin of Ly$\alpha$ by looking at the possibility of it originated from the Slug-DSFG itself due to recombination. Given the estimated SFRs we expect the intrinsic Ly$\alpha$ luminosity of Slug-DSFG to be $2.9\pm1.7\times10^{44}$\,erg s$^{-1}$, assuming a case B recombination ratio in luminosity of $L_{\rm Ly\alpha}$/$L_{\rm H\alpha}$=8.7 \citep{Henry:2015aa} and the conversion between H$\alpha$ and SFR provided in \citet{Kennicutt:2012aa}. Dust attenuation is estimated to be $A_{\rm V}=2.6$ based on the best-fit SED model, however a Bayesian analysis performed by {\sc cigale} shows $A_{\rm V} = 1.05\pm0.65$, suggesting a highly skewed probability distribution. If $A_{\rm V}=2.6$ is adopted, and assuming a Calzetti attenuation curve, the expected observed Ly$\alpha$ luminosity should have been about $2.5\pm1.5\times10^{41}$\,erg s$^{-1}$. If the probabilistic distribution is adopted for $A_{\rm V}$ the expected Ly$\alpha$ luminosity is instead $1.7^{+2.2}_{-1.4}\times10^{43}$\,erg s$^{-1}$. These values appear in agreement with each other due to large uncertainties. It therefore remains inconclusive regarding the origin of the Ly$\alpha$ emission from the Slug-DSFG, in particular whether it mostly emits from the Slug-DSFG via recombination, or due to scattering of other nearby sources from co-moving gas around the Slug-DSFG.

\subsection{The origin of the Slug ELAN}

\subsubsection{Brief overview}
Since the discovery of the Slug ELAN via narrow-band imaging \citep{Cantalupo:2014aa}, various follow-up observations have been made to study the origin of this exceptional Ly$\alpha$ structure which was initially interpreted as an intergalactic filament. \citet{Martin:2015aa} imaged the nebula with the Palomar Cosmic Web Imager (PCWI), which allows studies of kinematic properties of Ly$\alpha$ emission via integral field observations, and they found that the brightest Ly$\alpha$ emission region appears to be an extended rotating hydrogen disk with a diameter of about 125\,kpc in physical distance, possibly suggesting a cold flow disk similar to those predicted by some models \citep{Stewart:2011aa}. In the meantime, deep long-slit spectra taken from the Keck telescope were taken on various slit positions and angles, targeting multiple lines such as H$\alpha$, N\,{\sc ii}, He\,{\sc ii}, and C\,{\sc iv}, on both the nebula and a few selected nearby compact sources \citep{Arrigoni-Battaia:2015aa,Leibler:2018aa}. Through the measured line ratios between Ly$\alpha$ and H$\alpha$, as well as those between Ly$\alpha$ and He\,{\sc ii}, it has been suggested that the nebula is highly ionized, and has high gas densities ($n>1$\,cm$^{-3}$) and large clumping factors ($C\sim1000$). In addition, with these higher velocity resolution observations they were able to resolve the kinematic structures into a few distinct components, suggesting a more complex dynamical nature of the nebula than that suggested by the results from \citet{Martin:2015aa}. The hypothesis of having distinct kinematic components has recently found support from deep MUSE observations where sharp variations of Ly$\alpha$-to-He\,{\sc ii} line ratios have been reported \citep{Cantalupo:2019aa}. Understanding each of these kinematic components would help capture the full picture regarding the origin of the Slug ELAN. 

{\ As reviewed in the \autoref{sec:intro} and now more clearly shown} in the middle panel of \autoref{fig:fig5}, there appears to be three main kinematic structures; If referencing to UM287, the first is found to be at about $400$\,km\,s$^{-1}$, dubbed region c in \citet{Cantalupo:2019aa}, consisting of source 'C' and 'D' as well as the He\,{\sc ii} extended emission. The second component is at about $-500$\,km\,s$^{-1}$, corresponding to the 'bright tail' of the Ly$\alpha$ nebula. And finally the third one can be found at about $-300$\,km\,s$^{-1}$, which was dubbed region 1 by \citet{Leibler:2018aa}.

\subsubsection{Revisiting the bright tail}\label{sec:brightail}
In addition to the Slug-DSFG, our new $J$-band image has yielded a new continuum detection {\ (RA (J2000): 00:52:02.90 Dec(J2000):+01:01:23.85)} that may be part of the 'bright tail'. As shown in \autoref{fig:fig6}, the newly detected $J$-band continuum is marked by black crosses, and it is located at exactly the end of the bright tail where there are also $V$-band continuum emissions. The measured $J$-band flux density is 22.05$\pm$0.05 {\ (signal-to-noise ratio of 20)} in AB magnitude. Given the number density of our $J$-band image, the probability of finding a source brighter than this flux density by chance within a 1$''$ radius circular area centered at the detected location, roughly the size of the end of the bright tail, is about 1\%. While we do not have redshift measurements to confirm its association to the ELAN, the low probability of a random chance projection means that this $J$-band source could be part of the bright tail. 

Previously, the bright tail has been proposed to be a gas component that sits much further away from UM287 compared to regions 'c' \citep{Cantalupo:2019aa}. {\ This is based on the fact that the measured Ly$\alpha$ luminosity is much fainter than expected if instead it is the much higher gas density, not further distances, that drives the low He\,{\sc ii}-to-Ly$\alpha$ ratio.} However, together with the $V$-band continuum the newly discovered $J$-band continuum source could complicate the scenario. First the existence of a m$_J$=22 continuum source suggests that it is likely a galaxy, like the $V$-band continuum source {\ (RA (J2000): 00:52:02.81 Dec(J2000):+01:01:25.79)}, with a ISM type of density so much higher than the one typically assumed for halo gas, which may additionally serve to obscure Ly$\alpha$ and He\,{\sc ii}. Indeed, the fact that the southern edges of He\,{\sc ii} emission appearing to go around the two continuum sources could suggest that they are attenuated by these two sources that are possibly sitting in between the line-of-sight to region 'c'. With the possibility of galaxies being present in the bright tail and causing obscuration, one may need to re-evaluate all the possible factors, including distances, gas densities, and scattering effects, considering the observed Ly$\alpha$ luminosity and HeII-to-Ly$\alpha$ ratios, in order to understand its origin. Without knowing the exact redshifts of these two continuum sources we defer this exercise to future studies.

\begin{figure*}
	\begin{center}
		%\leavevmode
		\includegraphics[width=0.9 \textwidth]{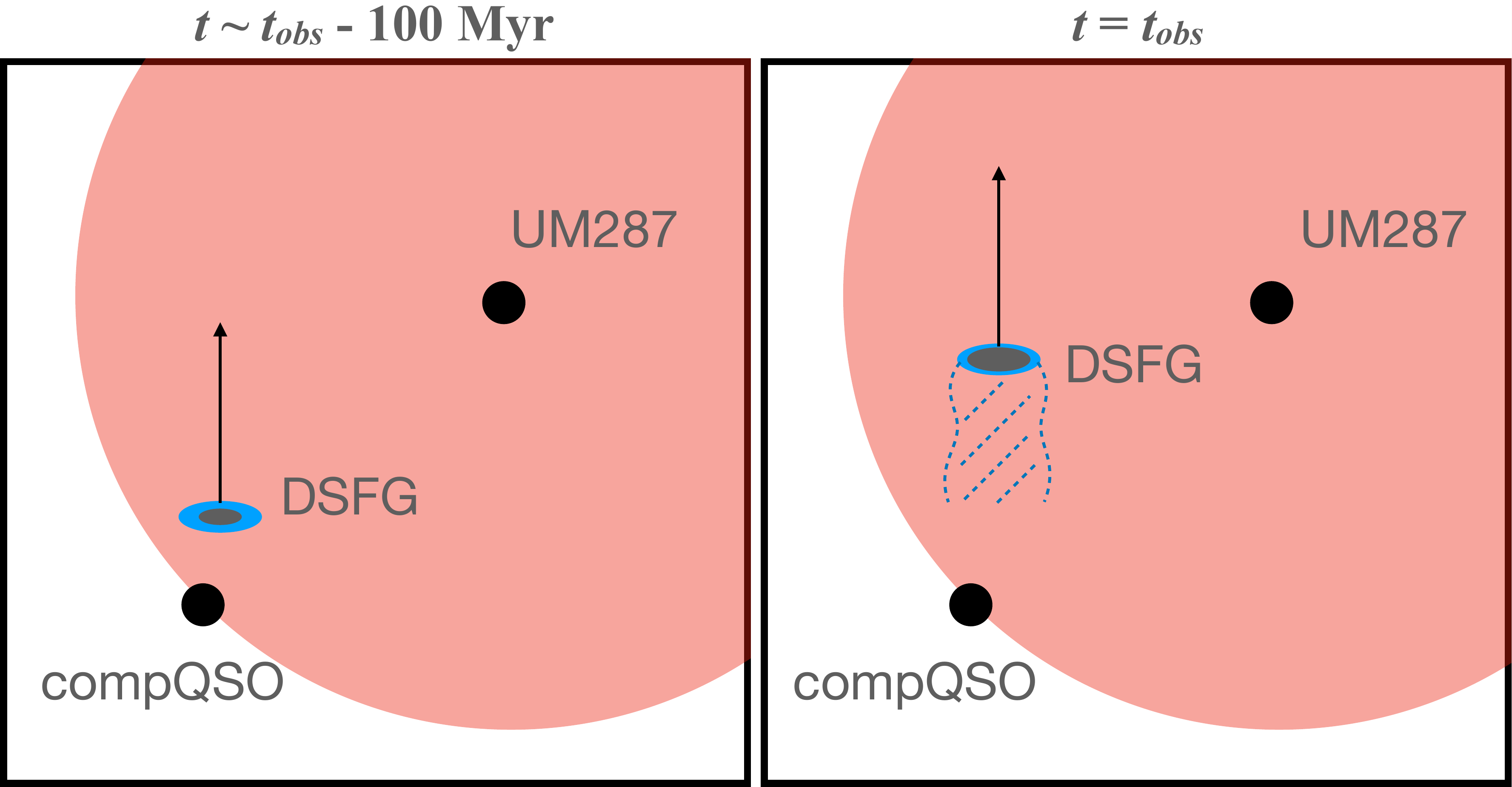}
		\caption{A schematic figure to show the proposed hypothetical scenario regarding the evolution of the newly discovered Slug-DSFG, where in the left shows the Slug-DSFG just entering the halo where gas stripping started to become effective. For simplicity we assume no relative movement between UM287 and the companion quasar. As shown in the right, we suspect that over the past $\sim$100\,Myr the Slug-DSFG has moved to the north by $\sim$100\,kpc in projected distance with a projected velocity of $\sim$1000\,km\,s$^{-1}$, in the meantime the stripped gas has been slowly entrained via shocks and gas mixing by the massive ($\sim$10$^{12}$-10$^{13}$\,M$_\odot$) hot halo of UM287 and a virial radius of $\sim$150\,kpc. The Slug-DSFG has also been experiencing strangulation, where the hot halo stops further accretion of cold gas so the gas reservoir has slowly drained out, therefore decreasing both the gas fraction and gas-to-dust ratios.}
		\label{fig:fig7}
	\end{center}
\end{figure*}

\subsubsection{Slug-DSFG and ram pressure stripping}
\label{sec:strip}
The newly discovered Slug-DSFG appears to have a velocity consistent with that of the Ly$\alpha$ component at $-300$\,km\,s$^{-1}$, the region 1 defined in \citet{Leibler:2018aa}. Indeed, the slit 2 of our Keck/LRIS spectrum shows that the Slug-DSFG sit at the far end of this Ly$\alpha$ filament (\autoref{fig:fig5}), suggesting that they belong to the same structure. The fact that we find possible evidence of gas stripping from the Slug-DSFG could further suggest that perhaps the Ly$\alpha$ filament is a result of this stripped gas. The smooth velocity gradient from $-300$\,km\,s$^{-1}$ to the systemic of UM287 could indicate that the stripped gas by ram pressure is gradually slowed down by the hot gas of the quasar halo as soon as it leaves the Slug-DSFG. The morphology of this stripped gas appears elongated to the north-south direction, with an end-to-end size of about 100\,kpc in projected distance. Together with the velocity gradient it could be that the Slug-DSFG have been moving from south to north in the sky while experiencing gas stripping. It is unclear what is the tangential peculiar velocity of the Slug-DSFG, but it would take about a crossing time of 100\,Myr to move across 100\,kpc assuming a projected velocity of about 1000\,km\,s$^{-1}$, which is roughly the escape velocity for halos similar to that of UM287 ($\sim$10$^{13}$\,M$_\odot$; e.g., \citealt{Miller:2018aa}).

The question is then can the stripped gas live for that long? The survivability of cold gas in shocked regions has been a challenging issue to models, mainly due to the fact that cold clouds are destroyed faster than their speed being equalized to the surrounding medium. However, recent theoretical studies tackling the acceleration of cold clouds by hot out-flowing winds have shown that efficient radiative cooling from the mixing gas can prolong the lifetime of clouds via mass accretion, such that the clouds can survive and even grown in mass until they are entrained \citep{Gronke:2018aa,Gronke:2020aa}. Based on these models, the time scale, dubbed the drag time $t_{\rm drag}$, that is required for the clouds with a size of $r_{\rm cl}$ to be accelerated to the velocity of the hot wind $v_{\rm wind}$, is $t_{\rm drag}\propto\chi r_{\rm cl}/v_{\rm wind}$, where $\chi$ is the overdensity of the cloud compared to the wind which is typically 100-1000.

Since the physical processes involved are similar, meaning acceleration (in our case de-acceleration) of cold clouds by surrounding hot gas via shocks and gas mixing, we assume these models to be applicable to the situation of ram pressure stripping discussed here. In such a case $v_{\rm wind}$ would become the peculiar velocity of the Slug-DSFG with respect to the quasar halo. Assuming again the peculiar velocity of 1000\,km\,s$^{-1}$, $\chi$=1000, and the size of the giant molecular clouds of $r_{\rm cl}=100$\,pc, as observed by lensed dusty galaxies \citep{Swinbank:2015aa,Dessauges-Zavadsky:2019aa}, $t_{\rm drag}$ is about 100\,Myr. While this remains a zeroth order estimate, as all assumed parameters are not well constrained, it is intriguing that with reasonable assumptions the drag time agrees with the crossing time of Slug-DSFG, supporting the hypothesis of the Ly$\alpha$ filament being the stripped gas slowed down by the hot halo gas.

Building on these arguments, we can also estimate the amount of gas mass stripped. Following \citet{Gunn:1972aa}, for a galaxy with a stellar surface density $\sigma_s$ and gas surface density $\sigma_g$ and moving with a peculiar velocity $v$ with respect to the hot halo of a volume density $\rho_{\rm halo}$, ram pressure stripping starts to become efficient  once $\rho_{\rm halo} v^2>2\pi G\sigma_s\sigma_g$. Hydro-dynamical simulations of 10$^{12}$\,M$_\odot$ halos at $z\sim2$, similar to the halo hosting UM287, have shown that $\rho_{\rm halo} \sim 10^{-3.5}$\,cm$^{-3}$ \citep{Nelson:2016bb} at about half of the virial radius, based on the projected distance between Slug-DSFG and UM287. For the mass profile of Slug-DSFG, following \autoref{sec:gf} we assume an exponential disk profile with a half-light radius of 3\,kpc and a gas fraction of 0.2. For a velocity of 1000\,km\,s$^{-1}$ we find that ram pressure from the hot halo could remove gas of the Slug-DSFG beyond a radius of 9\,kpc, and the gas masses stripped is about a few times $10^8$\,M$_\odot$, estimated by integrating the gas mass profile beyond this radius. This estimate likely lies in the lower end since the gas fraction was likely to be higher in the beginning when the Slug-DSFG first entered the hot halo, and the volume density $\rho_{\rm halo}$ could also be higher due to the limited resolution in simulations. If we increase the gas fraction to a typical SMG value of 0.5 and $\rho_{\rm halo}$ by one order of magnitude the stripped masses would be about 10$^9$\,M$_\odot$. 

As a consistency check we can compare this number to that estimated from the Ly$\alpha$ filament. Assuming a column density of ${\rm log}(N_{\rm H}/[{\rm cm^{-2}}])=20.5$, which is the median value from absorption studies of $z\sim2$ quasar halos \citep{Lau:2016aa}, and the area of $A=9.2$~arcsec$^2$ roughly defined by the ellipse in Figure~\ref{fig:fig6} (and consistent with the extend of emission along slit 2; Figure~\ref{fig:fig5}), we obtain a cool gas mass of $M_{\rm cool}^{\rm stripped}=A m_{\rm p}/X N_{\rm H}=2.2\times10^9$~M$_{\odot}$, where m$_{\rm p}$ is the proton mass and $X$=0.76 is the hydrogen mass fraction \citep{Pagel:1997aa}. Even though this value is again a zeroth order estimate, as the column densities inferred from absorption studies have a scatter of 1 dex and the area defined is rough, it is remarkably close to the estimates based on the Slug-DSFG. 

Interestingly, the amount of stripped gas turns out to be only a few percent of the total gas mass of the Slug-DSFG. So a more likely cause for a lower gas fraction and a lower gas-to-dust ratio for Slug-DSFG, or even UM287, is strangulation, where the gas was mainly consumed by star formation without replenishment from continuous feeding of cold gas. Given the estimated SFR from {\sc cigale} and the estimated crossing time of 100\,Myr, the gas masses consumed by star formation during this time would be about 2$\times$10$^{10}$\,M$_\odot$, accounting for most of deficit in gas masses when comparing to typical SMGs.

To provide a summary of what we have found from the Slug-DSFG, we sketch a schematic picture in \autoref{fig:fig7} to help visualize our proposed hypothesis. We suspect that the Slug-DSFG entered the halo about $\sim100$\,Myr ago when gas stripping started to become effective, and over the past $\sim100$\,Myr the Slug-DSFG has moved $\sim$100\,kpc to the north with a projected velocity of $\sim$1000\,km\,$^{-1}$, leaving behind $\sim$10$^8$-10$^9$\,M$_\odot$ stripped gas filaments that emit in Ly$\alpha$ (middle panel in \autoref{fig:fig5}) which become part of the ELAN and have been slowly entrained to the hot halo via shocks and gas mixing. The Slug-DSFG in the meantime has been experiencing strangulation where the hot halo stop further accretion of cold gas and the Slug-DSFG has consumed  $\sim$10$^{10}$\,M$_\odot$ via star formation, thus showing a deficit in gas fraction and gas-to-dust ratios (\autoref{sec:gf} and \autoref{sec:gdr}).

\section{Summary} \label{sec:sum}
To explore the origin of the enormous Ly$\alpha$ nebulae (ELAN), we have started A Multiwavelngth Study of ELAN Environment (AMUSE$^2$; \citealt{Arrigoni-Battaia:2018aa, Arrigoni-Battaia:2021aa, Nowotka:2021aa}) to measure and quantify the physical properties of sources that are located within or around ELAN. In this third paper of the series, we present single pointing band 4 ALMA CO(4-3) and [CI](1-0) observations toward an ELAN at $z=2.28$, dubbed the Slug ELAN \citep{Cantalupo:2014aa}. We summarize the results in the following.

\begin{enumerate}
\item We obtain significant detections in 2\,mm continuum from three sources and two of which also emit lines. Two of these three detected sources were known previously, UM287 and its companion quasar, and the other is a newly discovered source detected in both continuum and line. Our results are in favor of it being a dusty star-forming galaxy, dubbed Slug-DSFG, located approximately 100 projected kpc south-east of UM287 and sitting within the ELAN with a velocity difference of $-360\pm30$\,km\,s$^{-1}$ with respect to UM287.

\item With careful modeling of SED and dynamical analyses it is found that while their PDR conditions are similar to other comparison quasar and dusty galaxy samples at similar redshifts in the field, the Slug-DSFG and UM287 appear low in both gas fraction and gas-to-dust ratio, which have also been found on sources in another ELAN \citep{Arrigoni-Battaia:2021aa}, suggesting environmental effects from the host massive halos, with estimated masses of $\sim10^{13}$\,M$_\odot$ via various methods. 

\item Remarkably, our Keck long-slit spectra taken previously coincidentally covered the Slug-DSFG along with other regions of the nebula, revealing significant Ly$\alpha$ emission from the Slug-DSFG, as well as an extending Ly$\alpha$ tail toward the south with about 100\,kpc in length that connects both in real and velocity spaces with the Slug-DSFG.

\item With simple but realistic assumptions we propose that the Ly$\alpha$ tail is a result of stripped gas from the Slug-DSFG, which entered the influence of the hot halo $\sim$100\,Myr ago and has been moving from south to north on sky with a projected velocity of about 1000\,km\,s$^{-1}$. The gas mass stripped over this period of time is estimated to be about 10$^9$\,M$_\odot$, contributing to the dense cold gas reservoir that is believed to be responsible to the exceptional Ly$\alpha$ luminosity measured from this ELAN.

\item Under the proposed scenario, the estimated level of the stripped mass suggests that the main cause for the lower gas fraction and gas-to-dust ratio observed in the Slug-DSFG is more likely to be strangulation, where over the last $\sim$100\,Myr star formation has turned $\sim2\times10^{10}$\,M$_\odot$ of gas to stars without being further replenished, which is the amount comparable to the observed deficit. Assuming the current SFR remains largely unchanged and it is gravitationally bound to the halo, a natural prediction based on this proposed scenario is that the Slug-DSFG will run out of gas, thus quenched, in about $\sim$100\,Myr. 

\end{enumerate}
 
Our results highlight the importance of submillimeter observations of ELAN fields, which complement optical studies on properties of dust and molecular gas and help decipher the origin of ELAN. Last but not least, the estimated stripped gas mass is only about 10\% of the total dense and cool gas in the halo inferred from optical studies, which motivates deeper submillimeter observations in order to search for obscured but less massive satellites or companions.

\acknowledgments
{\ We thank the reviewer for a constructive feedback that has improved the manuscript.} We also thank Hiroyuki Hirashita, Zheng Cai, and Joe Hennawi for useful discussions. C.C.C. acknowledges support from the Ministry of Science and Technology of Taiwan (MOST 109-2112-M-001-016-MY3). This paper makes use of the following ALMA data: ADS/JAO.ALMA\#2018.1.00859.S. ALMA is a partnership of ESO (representing its member states), NSF (USA) and NINS (Japan), together with NRC (Canada), MOST and ASIAA (Taiwan), and KASI (Republic of Korea), in cooperation with the Republic of Chile. The Joint ALMA Observatory is operated by ESO, AUI/NRAO and NAOJ. The National Radio Astronomy Observatory is a facility of the National Science Foundation operated under cooperative agreement by Associated Universities, Inc.


\vspace{5mm}
\facilities{ALMA, VLT(MUSE, HAWK-I), Keck}


\software{astropy \citep{Astropy-Collaboration:2013aa},
        %   Cloudy \citep{2013RMxAA..49..137F}, 
        SExtractor \citep{Bertin:1996zr},
        UVMULTIFIT \citep{Marti-Vidal:2014aa}
          }



% \appendix

% \section{Appendix information}


\bibliography{bib}{}
% \bibliography{main}{}
\bibliographystyle{aasjournal}



% \listoftodos

\end{document}